\documentclass[reprint,
preprintnumbers,
nofootinbib,
amsmath,amssymb,
aps,
prd,
superscriptaddress,
]{revtex4-2}

\usepackage{graphicx}
\usepackage{dcolumn}
\usepackage{bm}
\usepackage[hidelinks]{hyperref}
\usepackage[all]{hypcap}
\usepackage{xcolor}
\usepackage{enumitem}
\usepackage{multirow}
\usepackage{orcidlink}
\usepackage{soul}

\hypersetup{
	colorlinks,
	linkcolor={red!50!black},
	citecolor={blue!50!black},
	urlcolor={blue!80!black}
}

\DeclareMathOperator{\Tr}{Tr}
\renewcommand{\Re}{\operatorname{Re}}

\begin{document}

\preprint{ADP-25-22/T1284}

\title{Center vortices in the novel phase of staggered fermions}

\author{Jackson A.\ Mickley\;\orcidlink{0000-0001-5294-2823}}
\affiliation{Center for the Subatomic Structure of Matter, Department of Physics, The University of Adelaide, South Australia 5005, Australia}

\author{Derek B.\ Leinweber\;\orcidlink{0000-0002-4745-6027}}
\affiliation{Center for the Subatomic Structure of Matter, Department of Physics, The University of Adelaide, South Australia 5005, Australia}

\author{Daniel Nogradi\;\orcidlink{0000-0002-3107-1958}}
\affiliation{E{\"o}tv{\"o}s Lor{\'a}nd University, Institute of Physics and Astronomy, Department of Theoretical Physics, Budapest, Hungary}

\begin{abstract}
The geometry of center vortices is studied in the novel lattice-artefact phase that appears with staggered fermions to elucidate any insight provided by the center-vortex degrees of freedom. For various numbers of fermion flavors, the single-site shift symmetry of the staggered-fermion action is broken in a finite region of the \((\beta, m)\) phase space. Simulations are performed with six degenerate fermion flavors and a range of \(\beta\) values that span the phase boundary. Center vortices are demonstrated to capture the broken shift symmetry that manifests in the unphysical phase. This persists at the level of each individual plaquette orientation, where it is revealed that only the plaquettes that span the broken dimension are affected. Several bulk center-vortex quantities, including the vortex and branching point densities, are considered to highlight other aspects of vortex geometry sensitive to the unphysical phase. A slight preference for the plaquettes affected by the broken shift symmetry to be pierced by a vortex is observed. This translates also to a greater branching point density in three-dimensional slices that span the broken dimension. Combined, these findings provide a novel characterization of the unphysical phase in terms of the fundamental center degrees of freedom.
\end{abstract}

\maketitle

\section{Introduction} \label{sec:intro}

The staggered fermion discretization on the lattice is advantageous due to its low computational cost and because it maintains a subset of chiral symmetry. Physical results, as with any other fermion formulation, are obtained in the continuum limit \(\beta\to\infty\), where \(\beta \sim 1/g_0^2\) is the bare coupling on the lattice. In the simplest cases at zero temperature, simulations are performed on some lattice volume, with bare masses \(m_i\) and coupling \(\beta\).

\begin{figure}
	\includegraphics[width=\linewidth]{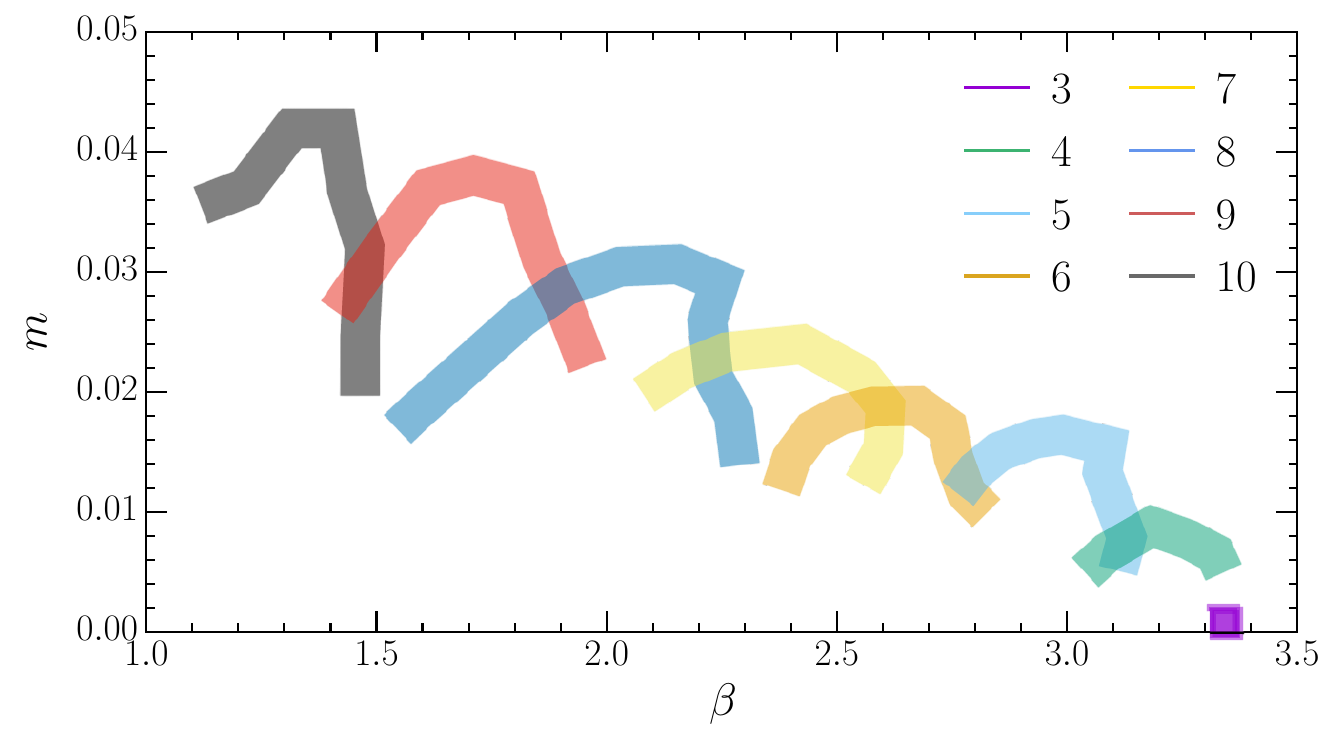}
	
	\vspace{-0.75em}
	
	\caption{\label{fig:phasediagram} Adapted from Fig.~2 in Ref.~\cite{Kotov:2021mgp}. Approximate locations of the broken-shift-symmetry phase boundaries in the \((\beta, m)\) plane for various numbers of fermion flavors \(N_f\). The phase is located under the curves and extends down to a bare quark mass \(m = 0\). The thickness of the curves reflects an uncertainty in the precise boundary locations. A transition does exist for \(N_f = 3\), but cannot be properly resolved at the shown scale.}
\end{figure}

The possibility of an unphysical lattice-artefact phase in the \((\beta, m)\) phase diagram was noted in Refs.~\cite{Lee:1999zxa, Aubin:2004dm}. Such a phase should naturally be avoided in taking the continuum limit. Concrete simulation results demonstrating the presence of the unphysical phase were reported in Refs.~\cite{Cheng:2011ic, Hasenfratz:2021zsl, Kotov:2021mgp} with various flavor content, all with gauge group \(\mathrm{SU}(3)\) and some amount of Dirac operator improvement in the form of smearing. Taking all quark flavors as degenerate, the unphysical phase corresponds to a finite region in the \((\beta,m)\) plane, as illustrated in Fig.~\ref{fig:phasediagram} for various flavor numbers. The unphysical phase has also been studied within the framework of chiral perturbation theory in Ref.~\cite{Aubin:2015dgk}. Interestingly, there is currently no convincing evidence for the presence of the unphysical phase with unimproved, unsmeared staggered fermions.

There is a symmetry breaking associated with the phase transition. In the unphysical phase the following single-site shift symmetry of the staggered fermion discretization \cite{Golterman:1984cy} is broken,
\begin{equation}
\begin{gathered}
\begin{aligned}
\chi(x) \to \xi_\mu(x)\, \chi(x+{\hat\mu})\,, && {\bar\chi}(x) \to {\bar\chi}(x+{\hat\mu})\, \xi_\mu(x)\,,
\end{aligned} \\
U_\mu(x) \to U_\mu(x+{\hat\mu})\;,
\end{gathered}
\end{equation}
while it is intact in the physical phase connected to the continuum limit. Here, \(\chi(x)\) is the staggered fermion field on the lattice with signs \(\xi_\mu(x) = (-1)^{\sum_{\nu > \mu} x_\nu}\), and \(U_\mu(x)\) are the usual link variables.

In this work, we examine this unphysical phase from the viewpoint of center vortices \cite{tHooft:1977nqb, tHooft:1979rtg, Nielsen:1979xu, Greensite:2003bk}. We aim to determine if center vortices can capture this physics and, if so, learn the insight provided through the relatively simple description of nonperturbative phenomena in QCD.

Extensive studies have demonstrated that the center-vortex degrees of freedom underpin many nonperturbative phenomena of lattice gauge theory, including confinement and chiral symmetry breaking \cite{DelDebbio:1996lih, DelDebbio:1997ep, Langfeld:1997jx, DelDebbio:1998luz, Faber:1997rp, Faber:1998qn, Kovacs:1998xm, Langfeld:1998cz, Engelhardt:1998wu, Bertle:1999tw, Engelhardt:1999fd, Engelhardt:1999wr, Faber:1999sq, deForcrand:1999our, deForcrand:2000pg, Kovacs:2000sy, Langfeld:2001cz, Langfeld:2003ev, Engelhardt:2003wm, Gattnar:2004gx, Bornyakov:2007fz, Bowman:2008qd, Bowman:2010zr, OMalley:2011aa, Hollwieser:2013xja, Hollwieser:2014soz, Trewartha:2015nna, Greensite:2016pfc, Trewartha:2017ive, Biddle:2022zgw, Biddle:2022acd, Mickley:2024zyg, Mickley:2024vkm}. In recent work, the dependence of center vortices on the number of flavors was analyzed in relation to the conformal window \cite{Mickley:2025mjj}, where care had to be taken to avoid the unphysical phase. It is thus of interest to investigate to what extent the vortex degrees of freedom capture the broken-shift-symmetry phase. If successful, significant insight into the unphysical phase may be obtained.

This paper is structured as follows. In Sec.~\ref{sec:centervortices}, we summarize the center-vortex approach and our lattice simulation details utilized for the study. The unphysical phase is demonstrated in Sec.~\ref{sec:unphysical}, initially for the untouched configurations before demonstrating that the broken shift symmetry is captured by the vortex-only fields. Section \ref{sec:statistics} presents a deep dive into several vortex properties that are sensitive to the phase transition. Finally, we summarize our findings in Sec.~\ref{sec:conclusion}.

\section{Center vortices} \label{sec:centervortices}
Center vortices~\cite{tHooft:1977nqb, tHooft:1979rtg, Nielsen:1979xu, Greensite:2003bk} are regions of the gauge field that carry magnetic flux quantized according to the center of SU(3),
\begin{equation}
	\mathbb{Z}_3 = \left\{ \exp\left(\frac{2\pi i}{3}\, n \right) \mathbb{I} \;\middle|\; n = -1,0,1 \right\} \,.
\end{equation}
Physical vortices in the QCD ground-state fields have a finite thickness. On the lattice, ``thin" center vortices are extracted through a gauge-fixing procedure that seeks to bring each link variable \(U_\mu(x)\) as close as possible to an element of \(\mathbb{Z}_3\), known as maximal center gauge (MCG). These thin vortices form closed surfaces in four-dimensional Euclidean spacetime, and thus one-dimensional structures in a three-dimensional slice of the four-dimensional spacetime.

Fixing to MCG is performed by finding the gauge transformation \(\Omega(x)\) to maximize the functional \cite{Montero:1999by}
\begin{equation}
	R = \sum_{x,\,\mu} \,\left| \Tr U_\mu^{\Omega}(x) \right|^2 \,.
\end{equation}
The links are subsequently projected onto the center,
\begin{equation}
	U_\mu^{\Omega}(x) \longrightarrow Z_\mu(x) = \exp\left(\frac{2\pi i}{3} \, n_\mu(x) \right) \mathbb{I} \in \mathbb{Z}_3 \,,
\end{equation}
with \(n_\mu(x) \in \{-1,0,1\}\) identified as the center phase nearest to \(\arg \Tr U_\mu(x)\) for each link. Finally, the locations of vortices are identified by nontrivial plaquettes in the center-projected field,
\begin{equation} \label{eq:centerprojplaq}
	P_{\mu\nu}(x) = \prod_\square Z_\mu(x) = \exp\left(\frac{2\pi i}{3} \, m_{\mu\nu}(x) \right)\mathbb{I}
\end{equation}
with \(m_{\mu\nu}(x) = \pm 1\). The value of \(m_{\mu\nu}(x)\) is referred to as the \textit{center charge} of the vortex, and we say the plaquette is pierced by a vortex.

Although gauge dependent, numerical evidence indicates that the projected vortex locations correspond to the physical ``guiding centers" of thick vortices in the original fields~\cite{DelDebbio:1998luz, Langfeld:2003ev, Montero:1999by, Faber:1999gu}. This allows one to investigate the significance of center vortices through the vortex-only field \(Z_\mu(x)\).

To investigate center vortices in the unphysical phase, six degenerate flavors of staggered fermions are employed with a bare mass \(m = 0.015\) and \(\beta\) values ranging from \(2.5\)--\(2.9\). These parameters are seen from Fig.~\ref{fig:phasediagram} to span the boundary of the broken-shift-symmetry phase for \(N_f = 6\). Symmetric \(16^4\) lattice volumes are utilized over the full \(\beta\) range, with \(24^4\) volumes also considered for a subset of \(\beta\) values to demonstrate that our results are not afflicted by finite volume effects.

The ensembles are generated using the HMC \cite{Duane:1987de} and RHMC \cite{Clark:2006fx} algorithms, with two steps of stout-smeared links \cite{Morningstar:2003gk, BMW:2010skj} at smearing parameter \(\rho = 0.12\) in the fermion sector and the tree-level Symanzik-improved action \cite{Symanzik:1983dc, Symanzik:1983gh} in the gauge sector. This setup is identical to Refs.~\cite{Nogradi:2019iek, Nogradi:2019auv, Kotov:2021mgp}.

\section{Unphysical phase} \label{sec:unphysical}
The first point of consideration is to demonstrate the broken shift symmetry on our untouched ensembles, and whether this carries down to the projected vortex-only fields. The typical order parameter used to quantify the unphysical phase is \cite{Cheng:2011ic}
\begin{gather}
	\label{eq:Delta_muP} \Delta_\mu P = \langle P(x) - P(x+\hat{\mu}) \rangle_{x_\mu\,\mathrm{even}}\,, \\
	\label{eq:plaq_sum} P(x) = \sum_{\mu<\nu} \frac{1}{3} \Re\Tr P_{\mu\nu}(x) \,,
\end{gather}
where \(P_{\mu\nu}(x) = U_\mu(x) \, U_\nu(x+\hat{\mu}) \, U_\mu^\dagger(x+\hat{\nu}) \, U_\nu^\dagger(x)\), and for each \(\mu\) the average in Eq.~(\ref{eq:Delta_muP}) is taken only over even lattice sites in the \(\mu\) direction. The quantity \(P(x)\) is a sum over all plaquette orientations at \(x\). In the physical phase, we expect \(\Delta_\mu P = 0\) for each \(\mu\), with the understanding that there is no preferred sign for the difference \(P(x) - P(x+\hat{\mu})\) between adjacent plaquettes. In the unphysical phase, however, \(\Delta_\mu P\) develops a nonzero value along one or more dimensions.

\begin{figure}
	\includegraphics[width=\linewidth]{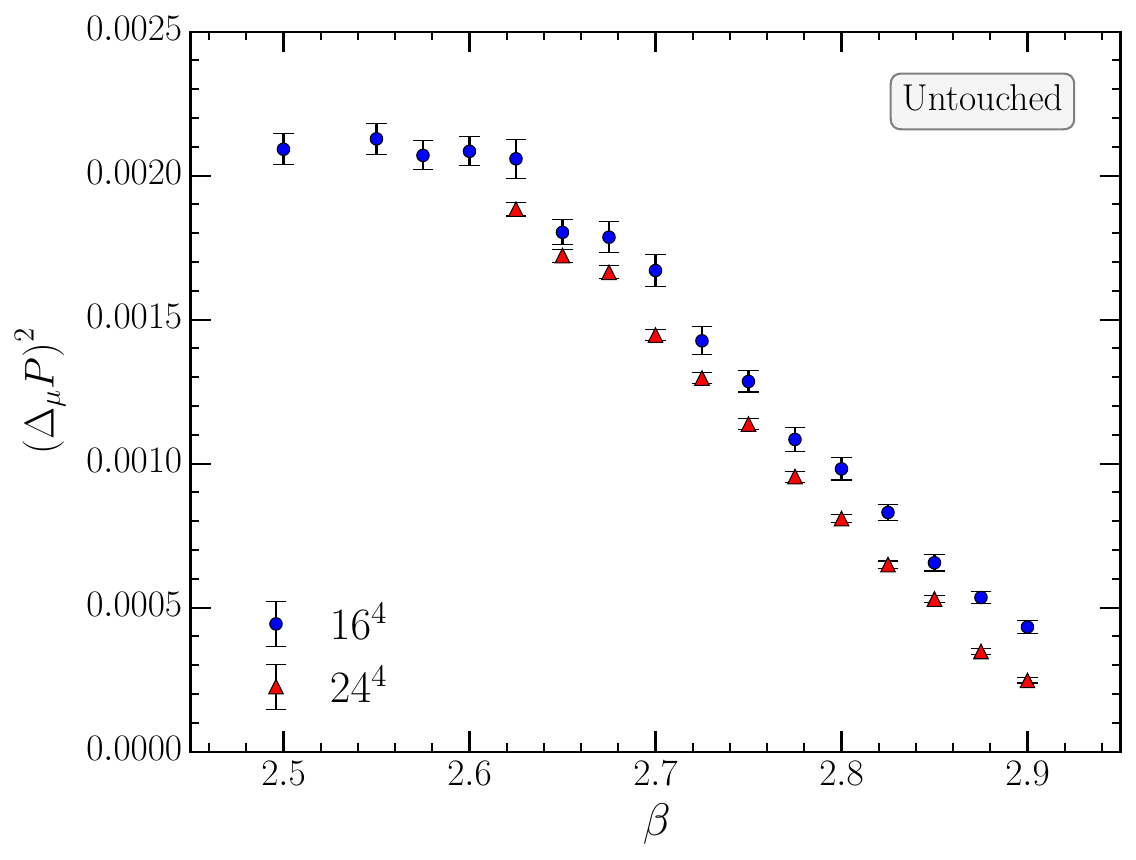}
	
	\vspace{-0.75em}
	
	\caption{\label{fig:Delta_mu_squared_untouched} The square \(\left(\Delta_\mu P\right)^2 = \left(\Delta_\mu P\right) \left(\Delta_\mu P\right)\) of the order parameter defined in Eq.~(\ref{eq:Delta_muP}) for our untouched configurations. It starts off close to zero at high \(\beta\) near the boundary of the unphysical phase, and steadily increases as \(\beta\) is decreased and we enter the broken phase proper.}
	
	\vspace{1.5em}
	
	\includegraphics[width=\linewidth]{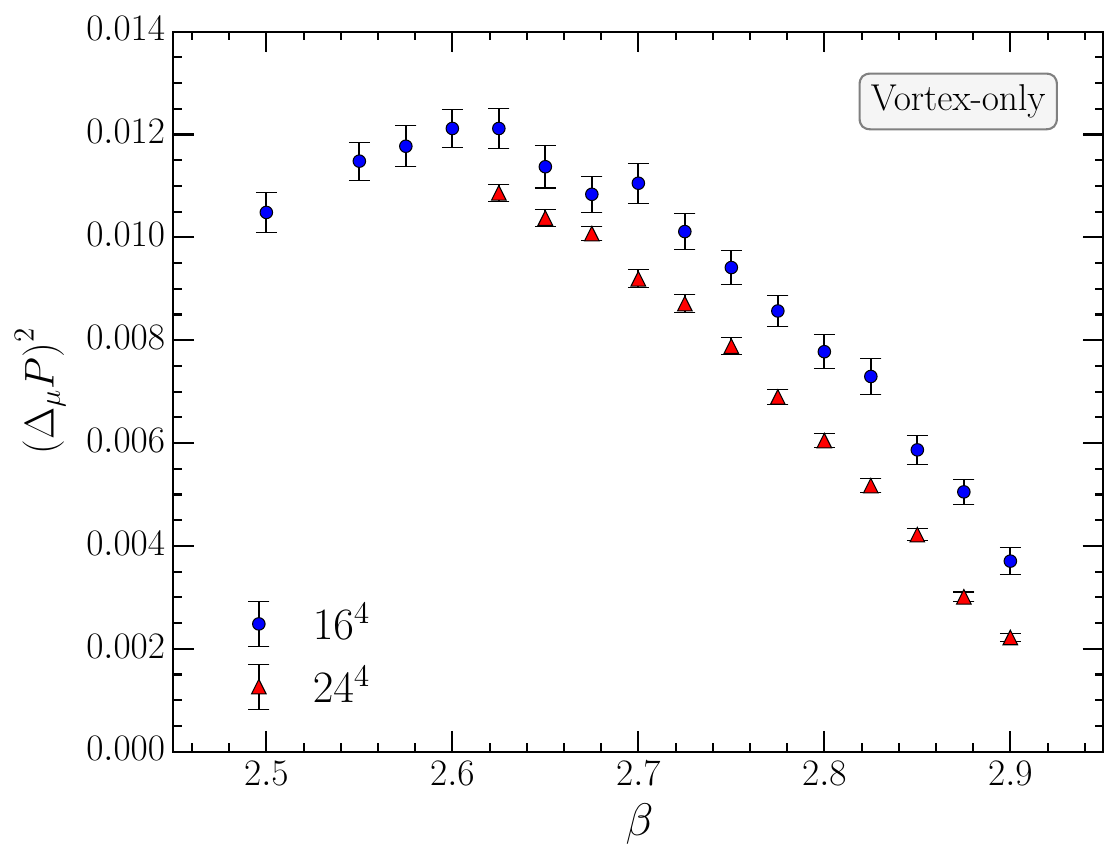}
	
	\vspace{-0.75em}
	
	\caption{\label{fig:Delta_mu_squared_vortex} Same as Fig.~\ref{fig:Delta_mu_squared_untouched} but for our vortex-only configurations, after fixing to maximal center gauge and projecting. A similar trend is observed as for the untouched configurations, though the order parameter is nearly an order of magnitude larger. Crucially, center vortices are seen to capture the broken shift symmetry.}
\end{figure}

To start, we collapse the dimensional information into a single quantity by calculating the square \(\left(\Delta_\mu P\right) \left(\Delta_\mu P\right)\) for each of our ensembles. Its evolution with $\beta$ is shown in Figs.~\ref{fig:Delta_mu_squared_untouched} and \ref{fig:Delta_mu_squared_vortex} for the untouched and vortex-only configurations, respectively.

Looking first at the untouched configurations, we find that \(\left(\Delta_\mu P\right)^2\) attains near-zero values in the upper end of our \(\beta\) range (\(\beta = 2.9\)) and gradually increases as \(\beta\) decreases. This is the expected outcome. The fact that \(\Delta_\mu P\) is not perfectly zero at our largest \(\beta\), in the physical phase, is a byproduct of the finite volume. One would expect this to approach zero as the infinite volume limit is taken. Indeed, a comparison of our lattice volumes reveals that \(\Delta_\mu P\) tends to be smaller on the larger \(24^4\) volume, particularly for higher \(\beta\). However, the smaller \(16^4\) volume is sufficient to highlight the transition into the broken-shift-symmetry phase.

Turning to the vortex-only fields (Fig.~\ref{fig:Delta_mu_squared_vortex}), we observe similar qualitative behavior in \(\left(\Delta_\mu P\right)^2\). Namely, it starts off small at large \(\beta\) and increases towards the left. This immediately indicates that the center-vortex degrees of freedom are sufficient to capture the broken shift symmetry. It is interesting to note that the value of \(\left(\Delta_\mu P\right)^2\) is approximately an order of magnitude larger on the vortex-only fields compared to the untouched configurations. This has a simple explanation. In the vortex fields, the difference between adjacent plaquettes is only nonzero when one plaquette is pierced by a vortex while the other is trivial. Owing to the discrete nature of the center, the smallest possible such difference is therefore \(\pm 1.5\). This is in contrast to the untouched configurations where the plaquette takes a continuous spectrum, and as such the difference $P(x) - P(x+\hat{\mu})$ tends to be smaller.

Having shown at a high level that center vortices are sensitive to the unphysical phase, we now wish to consider the broken symmetry in greater detail. It was mentioned that \(\Delta_\mu P\) can be nonzero in one or more dimensions at a time, and additionally this dimension can vary as a function of simulation time (i.e.\ on a configuration-to-configuration basis).

\begin{figure}
	\includegraphics[width=\linewidth]{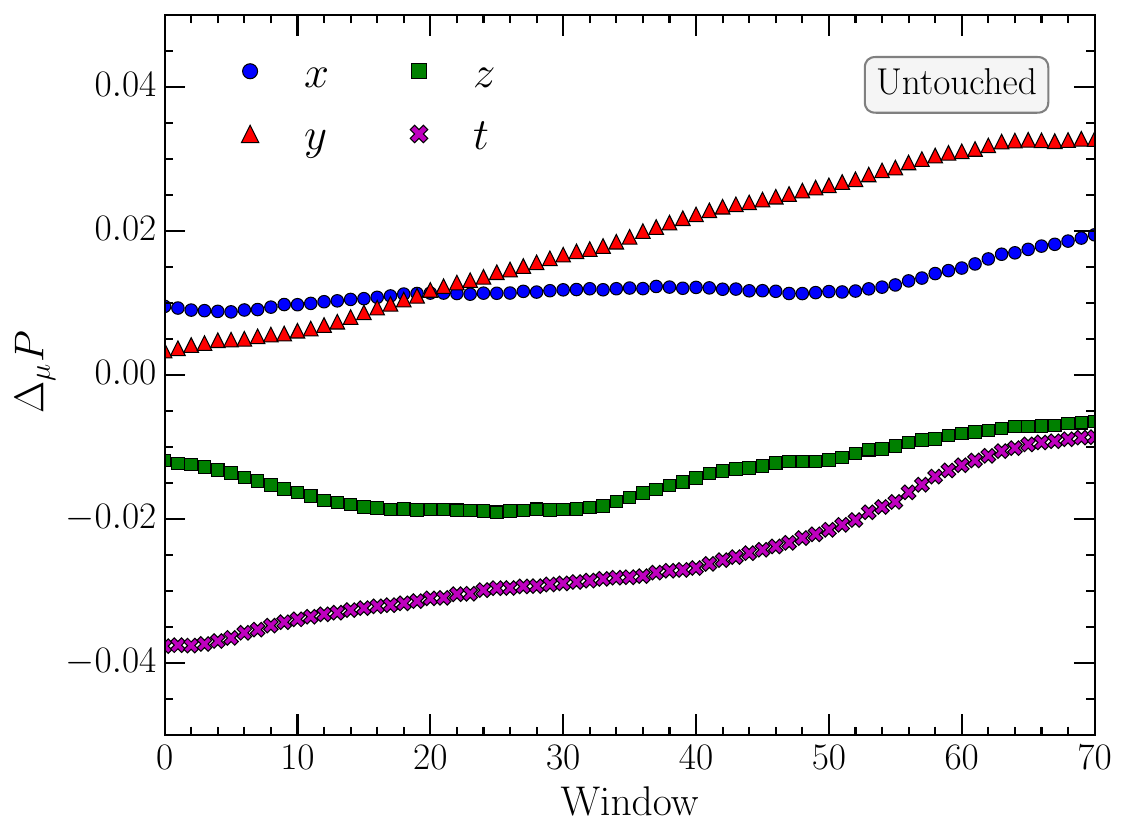}
	
	\vspace{-0.75em}
	
	\caption{\label{fig:Delta_mu_moving_average_untouched} A moving average of \(\Delta_\mu P\) over a window of 30 configurations in simulation time for each dimension \(\mu\), calculated on the untouched fields. This demonstrates how the shift symmetry can be broken along multiple dimensions simultaneously, and additionally that the strength of the symmetry breaking can vary with simulation time.}
	
	\vspace{1.5em}
	
	\includegraphics[width=\linewidth]{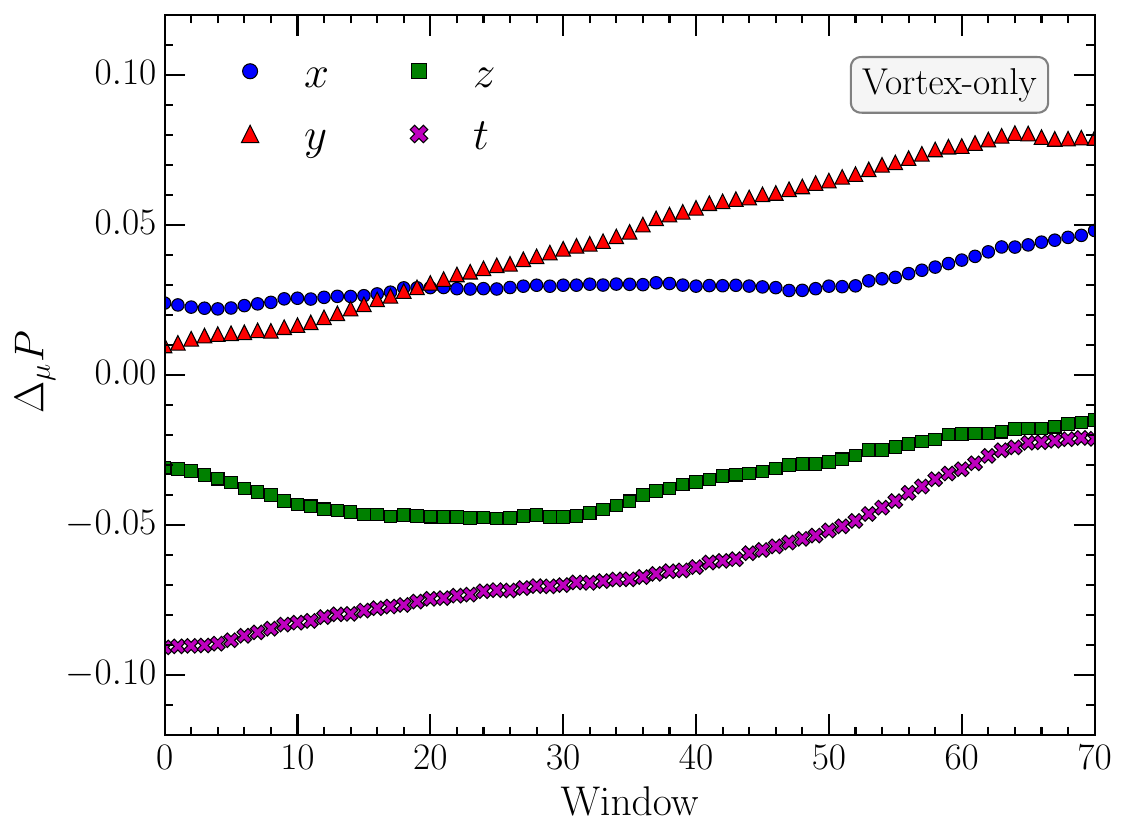}
	
	\vspace{-0.75em}
	
	\caption{\label{fig:Delta_mu_moving_average_vortex} Same as Fig.~\ref{fig:Delta_mu_moving_average_untouched} but for our vortex-only configurations.}
\end{figure}

To examine this, we select one \(24^4\) ensemble in the unphysical phase at \(\beta = 2.675\), and calculate \(\Delta_\mu P\) for each \(\mu\) under a moving average of 30 configurations in the HMC trajectory. This window size is chosen to be sufficiently large to produce a smooth curve as the window is shifted, while also being sufficiently small to demonstrate the variation in \(\Delta_\mu P\) with simulation time. To be precise, the value associated with the \(i\)th window of 30 configurations is calculated by averaging over configurations \(i\) through \(i + 29\) in simulation time, with ten HMC trajectories between each sampled configuration. The resulting evolution is presented for the untouched configurations in Fig.~\ref{fig:Delta_mu_moving_average_untouched}, and again for the vortex-only fields in Fig.~\ref{fig:Delta_mu_moving_average_vortex}.

Initially, the symmetry breaking is most pronounced in the temporal dimension \((\mu = t)\), for which the difference \(\Delta_\mu P\) is negative. This is seen to weaken as the 30-configuration window is shifted, while simultaneously becoming stronger in the \(y\) dimension where the difference is positive. This illustrates how the symmetry breaking can move between the dimensions, and is not locked into a single sign. Furthermore, there is a mild breaking observed in the \(x\) and \(z\) dimensions over a majority of the analyzed range. Nearly identical behavior is exhibited on both the untouched and vortex-only configurations, though the value of \(\Delta_\mu P\) is again considerably larger on the latter.

The gradual change in direction of the order parameter throughout the simulation is typical of spontaneous symmetry breaking in a finite volume, such as is also observed with the chiral condensate. This is in contrast to explicit symmetry breaking, which would fix the direction of the order parameter, though since no explicit breaking is present \(\Delta_\mu P\) is found to slowly rotate around the various dimensions \(\mu\).

In fact, it is possible to further decompose \(\Delta_\mu P\) down to the individual-plaquette level. So far, all plaquette orientations have been summed over in calculating \(\Delta_\mu P\), as described by Eq.~(\ref{eq:plaq_sum}). We are now interested in ascertaining whether \textit{all} plaquette orientations are affected if the shift symmetry is broken along a specific dimension \(\mu\), or only a subset. For instance, one might hypothesize that it is only those plaquettes that include the broken dimension \(\mu\) that are affected. This is the first time that the broken symmetry has been studied at this level.

\begin{figure}
	\includegraphics[width=\linewidth]{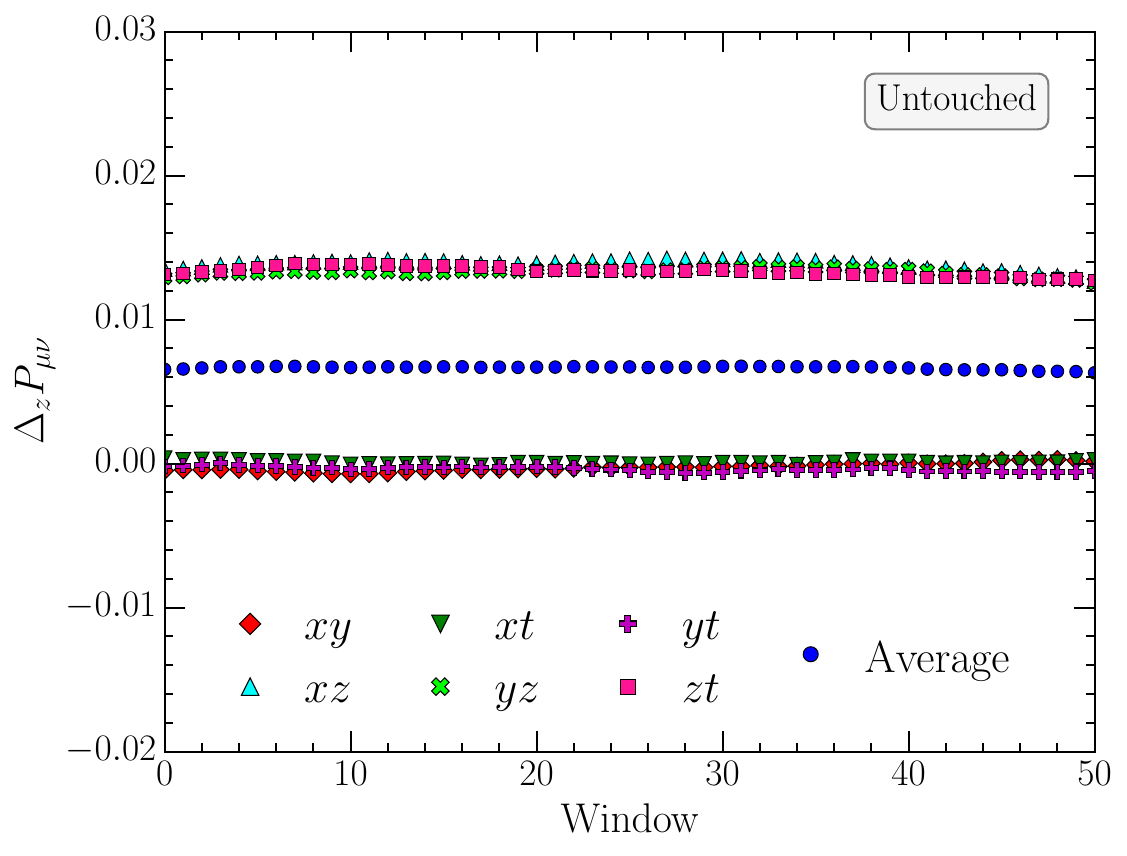}
	
	\vspace{-0.75em}
	
	\caption{\label{fig:Delta_mu_plaquettes_untouched} The order parameter \(\Delta_z P_{\mu\nu}\) for each distinct plaquette orientation \(\mu < \nu\) on our ensemble at \(\beta = 2.55\). A moving average is again utilized to show constancy with simulation time. This shows that it is only those plaquettes that include the shifted dimension (here, the \(z\) dimension) for which the symmetry is broken.}
	
	\vspace{1.5em}
	
	\includegraphics[width=\linewidth]{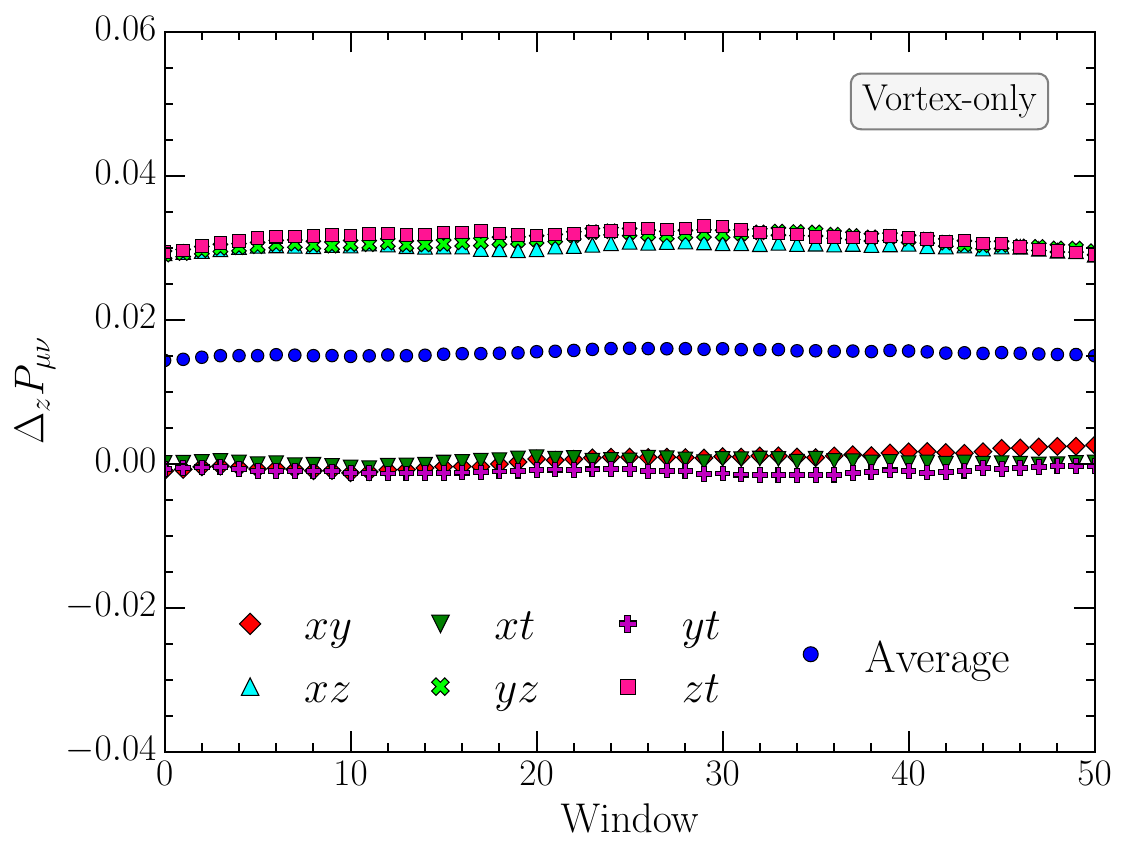}
	
	\vspace{-0.75em}
	
	\caption{\label{fig:Delta_mu_plaquettes_vortex} Same as Fig.~\ref{fig:Delta_mu_plaquettes_untouched} but for our vortex-only configurations.}
\end{figure}

To conduct this investigation, it is beneficial to choose an ensemble on which the symmetry is broken along one dimension over an extended simulation time. We are fortunate that such an example is provided by our ensemble at \(\beta = 2.55\), which is seen from Figs.~\ref{fig:Delta_mu_squared_untouched} and \ref{fig:Delta_mu_squared_vortex} to be our \(\beta\) value with near-maximal symmetry breaking for both the untouched and vortex-only configurations. On this ensemble, the symmetry is strongly broken along the \(z\) dimension (i.e.\ \(\mu = z\)) for 100 configurations. Instead of summing over all distinct plaquette orientations, we therefore proceed to calculate \(\Delta_z P_{\mu\nu}\) separately for every \(\mu < \nu\). The results are shown in Figs.~\ref{fig:Delta_mu_plaquettes_untouched} and \ref{fig:Delta_mu_plaquettes_vortex}, with the average over all orientations displayed for reference.

Here, we discover that it is in fact only the plaquettes that include the dimension under consideration for which the shift symmetry is broken. The individual order parameters for \(xy\), \(xt\), and \(yt\) plaquettes, none of which include the \(z\) dimension, are consistent with zero over the entire considered range. On the other hand, the equivalent computations for \(xz\), \(yz\) and \(zt\) plaquettes, all of which cover the \(z\) dimension, reveal persistently nonzero values of \(\Delta_z P_{\mu\nu}\), indicating breaking of the shift symmetry. The order parameter obtained from averaging over all plaquette orientation unsurprisingly lies halfway between the two groups.

As with the previous quantities, this decomposition is valid on both the untouched and vortex-only configurations. This soundly establishes that center vortices capture the essential aspects of the unphysical broken-shift-symmetry phase.

It is prescient to discuss the explicit interpretation of the broken symmetry on vortex configurations. As previously mentioned, the difference between adjacent plaquettes is only nonzero when one plaquette of the pair is pierced by a vortex. As such, if the order parameter in Eq.~(\ref{eq:Delta_muP}) is nonzero after averaging over all lattice sites, this indicates a preferred parity for nontrivial plaquettes. To be precise, if a plaquette \(P_{\mu\nu}(x)\) is pierced, then its real part \(\Re\Tr P_{\mu\nu}(x)\) is negative. As such, a negative value for the order parameter \(\Delta_\mu P\) indicates a preference for even plaquettes in the \(\mu\) dimension to be pierced by a vortex, while a positive value implies a preference for odd plaquettes to be pierced.

Ideally, we would like this preference to be discernible in visualizations of center-vortex structure. Prior work \cite{Biddle:2019gke} has developed the tools necessary to visualize center vortices. A three-dimensional cross section of the full four-dimensional lattice is taken by holding one of the coordinates fixed. For each plaquette in the three-dimensional slice, this leaves one orthogonal direction that is used to identify the plaquette. A vortex is then rendered as an arrow existing on the dual lattice and piercing the associated nontrivial plaquette.

It would be fascinating to observe the asymmetry between even and odd plaquettes in such a visualization, though for this to be possible the preference needs to be sufficiently strong. As such, we first seek to explicitly quantify the asymmetry between even and odd plaquettes. This is conducted as follows. On a given configuration, for each dimension \(\mu\) we count the number of even \(N_\mathrm{even}(\mu)\) and odd \(N_\mathrm{odd}(\mu)\) plaquettes in the \(\mu\) dimension that are pierced by a vortex. This is equivalent to counting the number of times that the difference between adjacent plaquettes in Eq.~(\ref{eq:Delta_muP}) is negative or positive, though it has a more precise meaning for the vortex fields. We then compute the ratios
\begin{equation}
	\label{eq:ratios} R_\mu = \frac{N_\mathrm{odd}(\mu)}{N_\mathrm{even}(\mu)} \,,
\end{equation}
with the ratio defined to be \(>1\) when \(\Delta_\mu P\) is positive.

\begin{figure}
	\includegraphics[width=\linewidth]{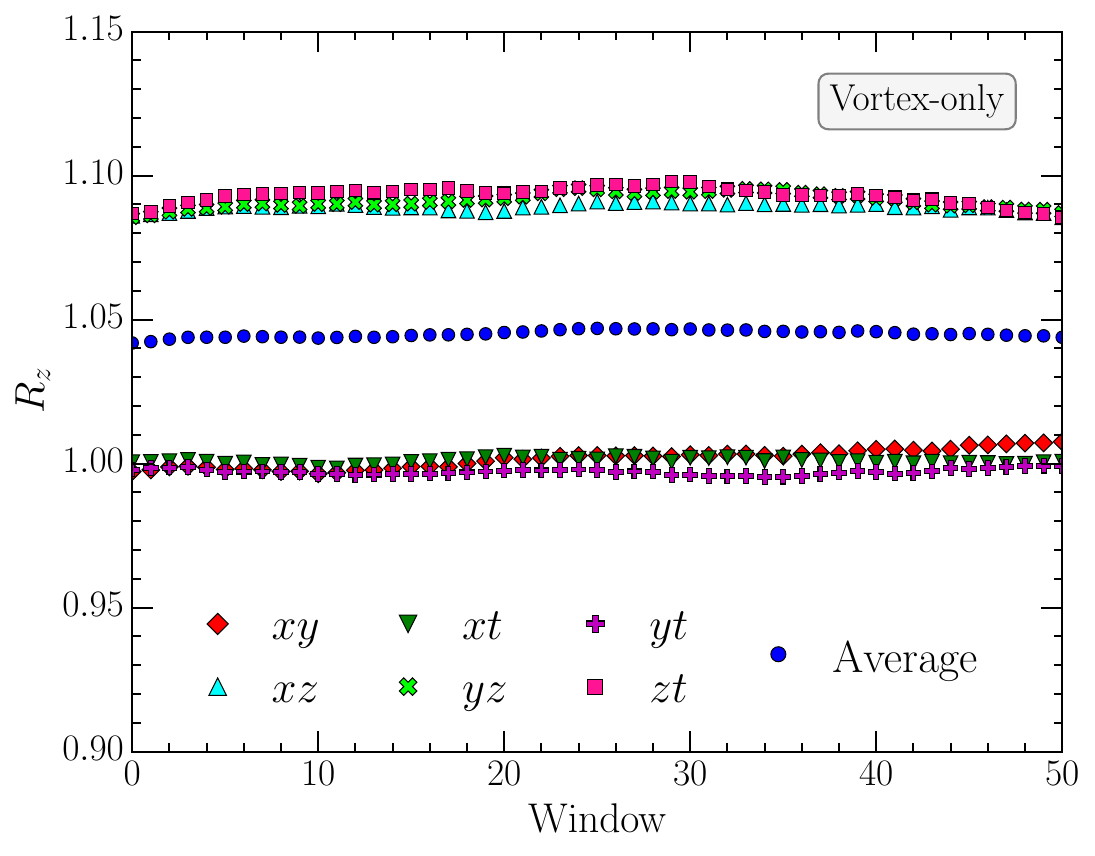}
	
	\vspace{-0.75em}
	
	\caption{\label{fig:ratio_plaquettes} The ratios \(R_z\) defined in Eq.~(\ref{eq:ratios}) for each distinct plaquette orientation on our ensemble at \(\beta = 2.55\). A moving average is utilized as in previous figures. The ratio is approximately unity for the plaquettes that exclude the \(z\) dimension, and considerably greater than unity for those that include the \(z\) dimension.}
\end{figure}

As with \(\Delta_\mu P\), these ratios can be calculated individually for each plaquette orientation. We exploit this to reproduce Fig.~\ref{fig:Delta_mu_plaquettes_vortex}, but for the ratio $R_z$ instead of the difference $\Delta_z P$. This is shown in Fig.~\ref{fig:ratio_plaquettes}.
Similar qualitative features are exhibited as in Fig.~\ref{fig:Delta_mu_plaquettes_vortex}. Namely, for the plaquette orientations that do not span the broken \(z\) dimension (\(xy\), \(xt\), \(yt\)), the ratio is approximately unity. This indicates there is no preference for even or odd plaquettes, and that the shift symmetry is intact. For the other plaquettes that do contain the \(z\) dimension, the ratio is notably greater than 1, a clear signal of the asymmetry.

However, the primary purpose of computing these ratios is to quantify the extent of the even-odd asymmetry. We see from Fig.~\ref{fig:ratio_plaquettes} that the maximum value attained within one window of configurations is \(\simeq 1.1\), signifying a 10\% asymmetry. That is to say, for every 10 even plaquettes pierced by a vortex, there are 11 odd plaquettes pierced. Unfortunately, it does not seem likely that such a discrepancy will be identifiable in visualizations. The maximum ratio on any single configuration was \(\simeq 1.15\), which, although slightly larger, still seems insufficient. Recalling that the ensemble used to produce these plots was chosen due to its maximal symmetry breaking, we therefore concede that the broken shift symmetry is not likely to be gleaned from visualizations.

\section{Vortex statistics} \label{sec:statistics}
Having verified that center vortices capture the essential broken shift symmetry, we now proceed to scrutinize whether any other aspects of vortex geometry are sensitive to the unphysical phase. This will include a consideration of bulk quantities such as the vortex and branching point densities, along with the local internal distribution of branching points throughout the vortex cluster.

\subsection{Vortex density} \label{subsec:vortexdensity}
\begin{figure}
	\includegraphics[width=\linewidth]{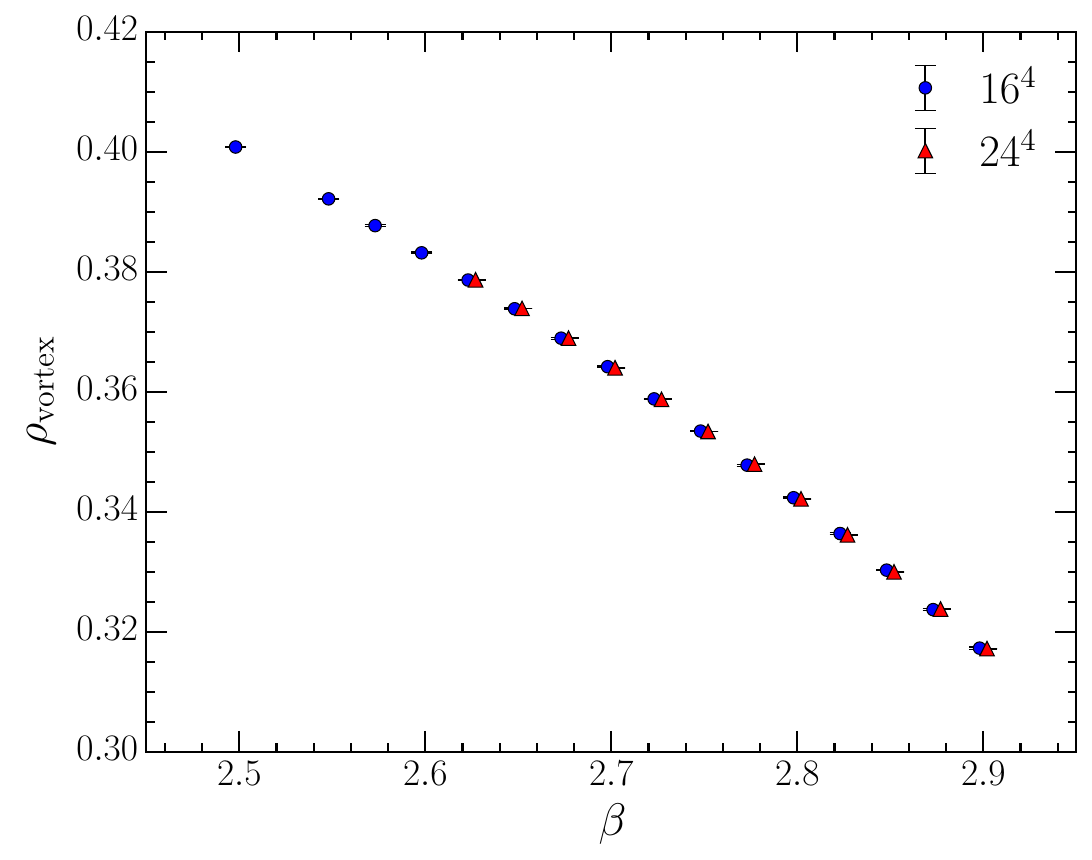}
	
	\vspace{-0.75em}
	
	\caption{\label{fig:vortexdensity} The dimensionless vortex density \(\rho_\mathrm{vortex}\) defined in Eq.~(\ref{eq:vortexdensity}) for each \(\beta\) value and volume. Bootstrap statistical uncertainties are displayed, though are too small (\(\sim \mathcal{O}(10^{-4})\)) to be resolved at this scale. There is a gentle increase in density towards lower \(\beta\) values, in line with an increasing ``roughness" in the gauge field as \(\beta\) decreases. No finite volume effects are seen to be present.}
\end{figure}

The first quantity we examine is the vortex density, defined simply as the total proportion of plaquettes pierced by a vortex,
\begin{equation} \label{eq:vortexdensity}
	\rho_\mathrm{vortex} = \frac{\mathrm{Number\ of\ nontrivial\ plaquettes}}{6\,N_\mathrm{sites}} \,,
\end{equation}
where \(N_\mathrm{sites}\) is the dimensionless lattice volume and \(\binom{4}{2} = 6\) counts the number of distinct plaquette orientations. As it stands in Eq.~(\ref{eq:vortexdensity}), \(\rho_\mathrm{vortex}\) is dimensionless; it is typically converted to a physical quantity through dividing by \(a^2\), the area of a single plaquette. However, the relevance of physical quantities such as the lattice spacing in the unphysical phase are doubtful, and as such we leave the vortex density dimensionless for all \(\beta\) values. Its behavior as a function of \(\beta\) is presented in Fig.~\ref{fig:vortexdensity}.

We note that the values obtained on \(16^4\) and \(24^4\) volumes are coincident with each other, corroborating the absence of finite volume artefacts. A soft increase in vortex density as \(\beta\) decreases is revealed. At first glance this may seem a product of entering the unphysical phase, though we believe the same shift in value would be observed irrespective of the phase's existence. Intuitively, smaller \(\beta\) values correspond to a ``rougher" gauge field, for which the natural consequence would be an increase in vortex matter. It appears likely that the trend in Fig.~\ref{fig:vortexdensity} is nothing but a reflection of this fact, and that \(\rho_\mathrm{vortex}\) would continue to climb even as one exits the unphysical phase again. Certainly, there is no rapid change to \(\rho_\mathrm{vortex}\) as the phase boundary is crossed. This contrasts the order parameter in Figs.~\ref{fig:Delta_mu_squared_untouched} and \ref{fig:Delta_mu_squared_vortex} that indicates we have entered a new phase.

\begin{figure}
	\includegraphics[width=\linewidth]{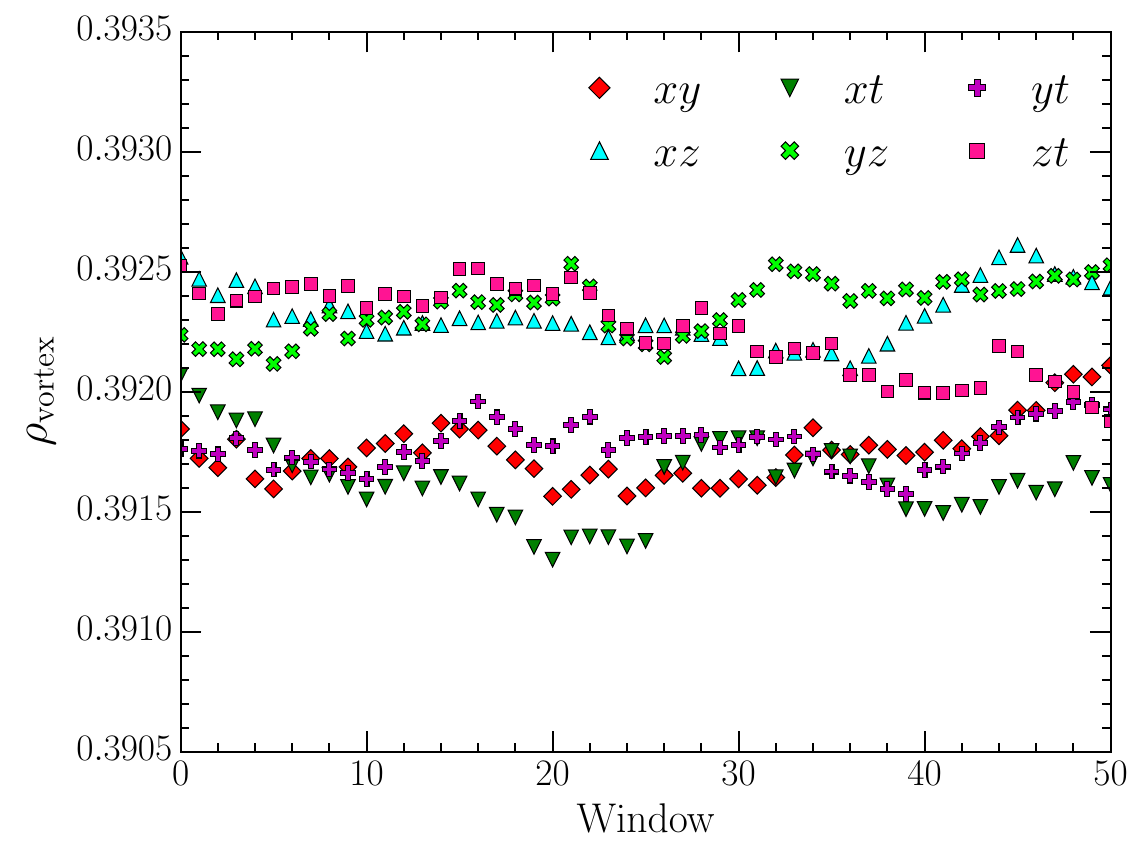}
	
	\vspace{-0.75em}
	
	\caption{\label{fig:plaquette_densities} The density of plaquettes pierced by a vortex for each plaquette orientation under a moving average for \(\beta = 2.55\). This hints at a subtle discrepancy in plaquettes pierced between those that span or do not span the broken dimension (in this case, the \(z\) dimension).}
	
	\vspace{1.5em}
	
	\includegraphics[width=\linewidth]{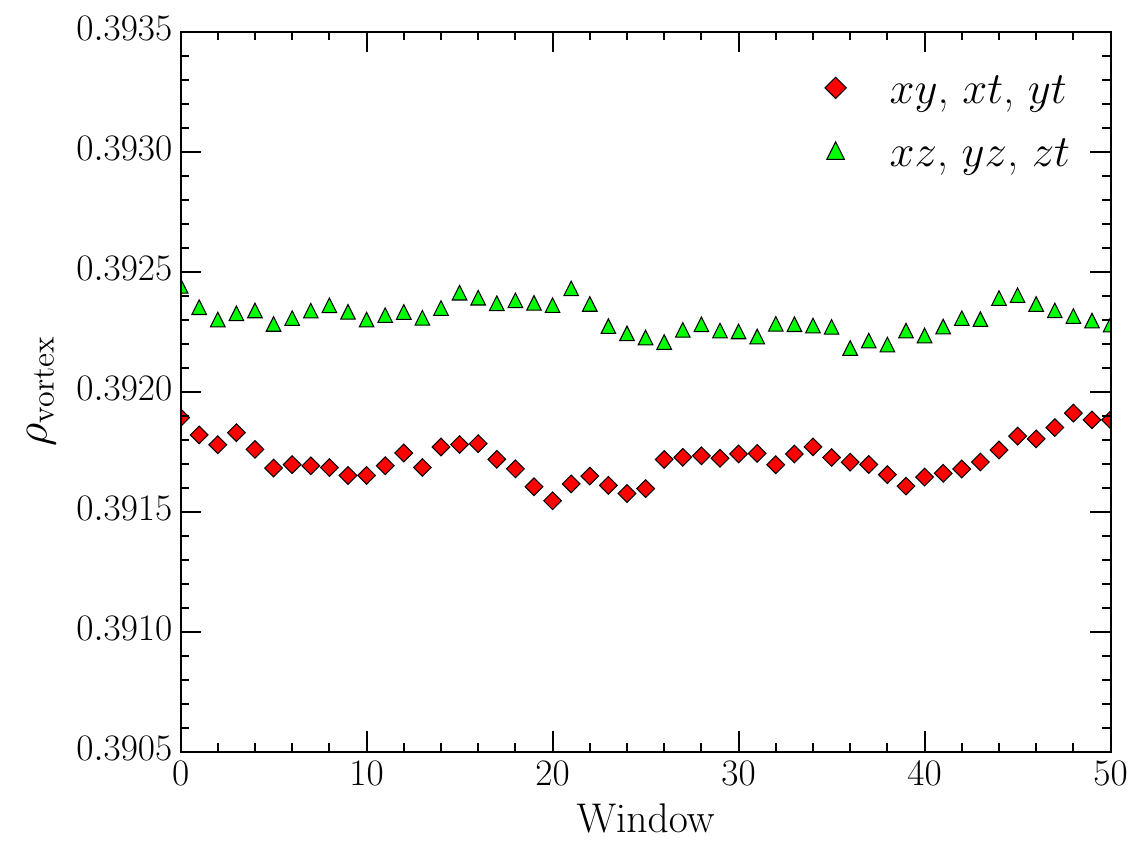}
	
	\vspace{-0.75em}
	
	\caption{\label{fig:plaquette_densities_averaged} Same as Fig.~\ref{fig:plaquette_densities} but averaged over the three plaquette orientations in each group.}
\end{figure}

Nevertheless, we discovered in Sec.~\ref{sec:unphysical} that the defining property of the unphysical phase, the broken shift symmetry, only impacts the plaquettes that span the broken dimension. Instead of considering a net density that incorporates all plaquettes, as in Eq.~(\ref{eq:vortexdensity}), it is thus prudent to decompose the vortex density into individual plaquette orientations. That is to say, we define a \(\rho_\mathrm{vortex}(\mu,\nu)\) for each \(\mu,\nu\) as the proportion of \(\mu\)-\(\nu\) plaquettes pierced. In the physical phase these are all equal, though it is possible a discrepancy will form in the unphysical phase. To investigate this, we again return to our \(\beta = 2.55\) ensemble on which we expect any difference to be maximal, and compute a moving average of the individual-plaquette densities. The result is shown in Fig.~\ref{fig:plaquette_densities}.

Remarkably, there is a suggestion of a discrepancy between the different plaquette orientations. That is, the \(xz\), \(yz\), and \(zt\) densities appear to sit above the \(xy\), \(xt\) and \(yt\) densities over the entire range. Recalling it is the \(z\) dimension along which the symmetry is broken, this is precisely the manner in which a difference might appear. To aid in verifying whether this is genuine, we reproduce the plot in Fig.~\ref{fig:plaquette_densities_averaged} averaging over the three plaquette orientations within each group. This smooths out some of the noise present in Fig.~\ref{fig:plaquette_densities}, and reveals clearly a greater density of pierced plaquettes containing the broken \(z\) dimension. Furthermore, the split persists over all 100 configurations; this is the purpose of continuing to utilize a moving average.

We have performed tests at other \(\beta\) values and verified that this statement holds true when the shift symmetry is broken along other dimensions, and also when the order parameter \(\Delta_\mu P\) is negative instead of positive. We have therefore discovered that there is a slight preference for the plaquettes that include the broken dimension to be pierced by a vortex. It is not immediately clear why the plaquettes that exhibit the broken shift symmetry should be pierced slightly more often, though we note the effect is small enough that it is unlikely to compromise any conclusions drawn from studying the total vortex density.

\subsection{Branching point density} \label{subsec:branchingpoints}
\begin{figure}
	\centering
	\includegraphics[width=0.49\linewidth]{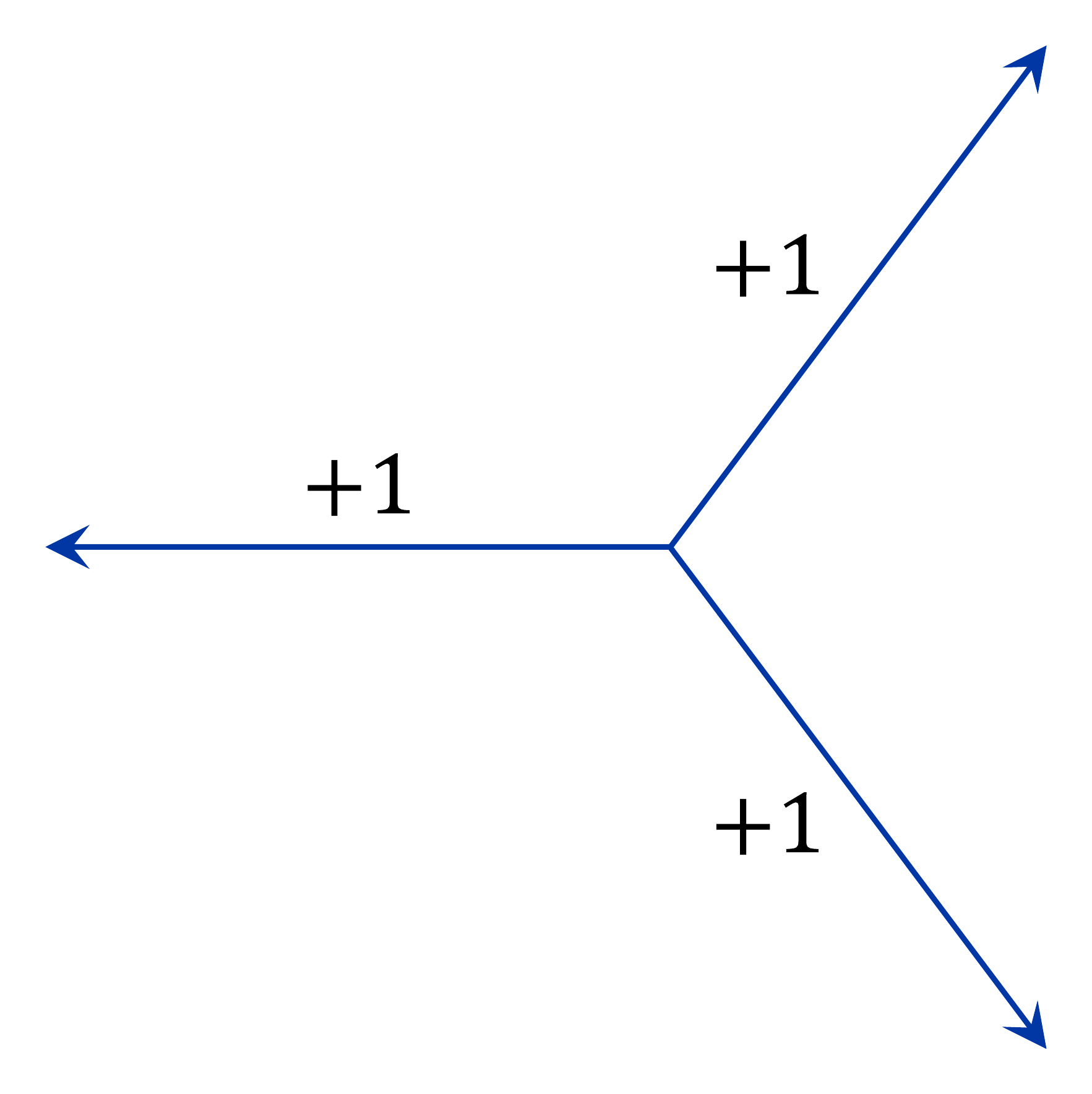}
	\hfill
	\includegraphics[width=0.49\linewidth]{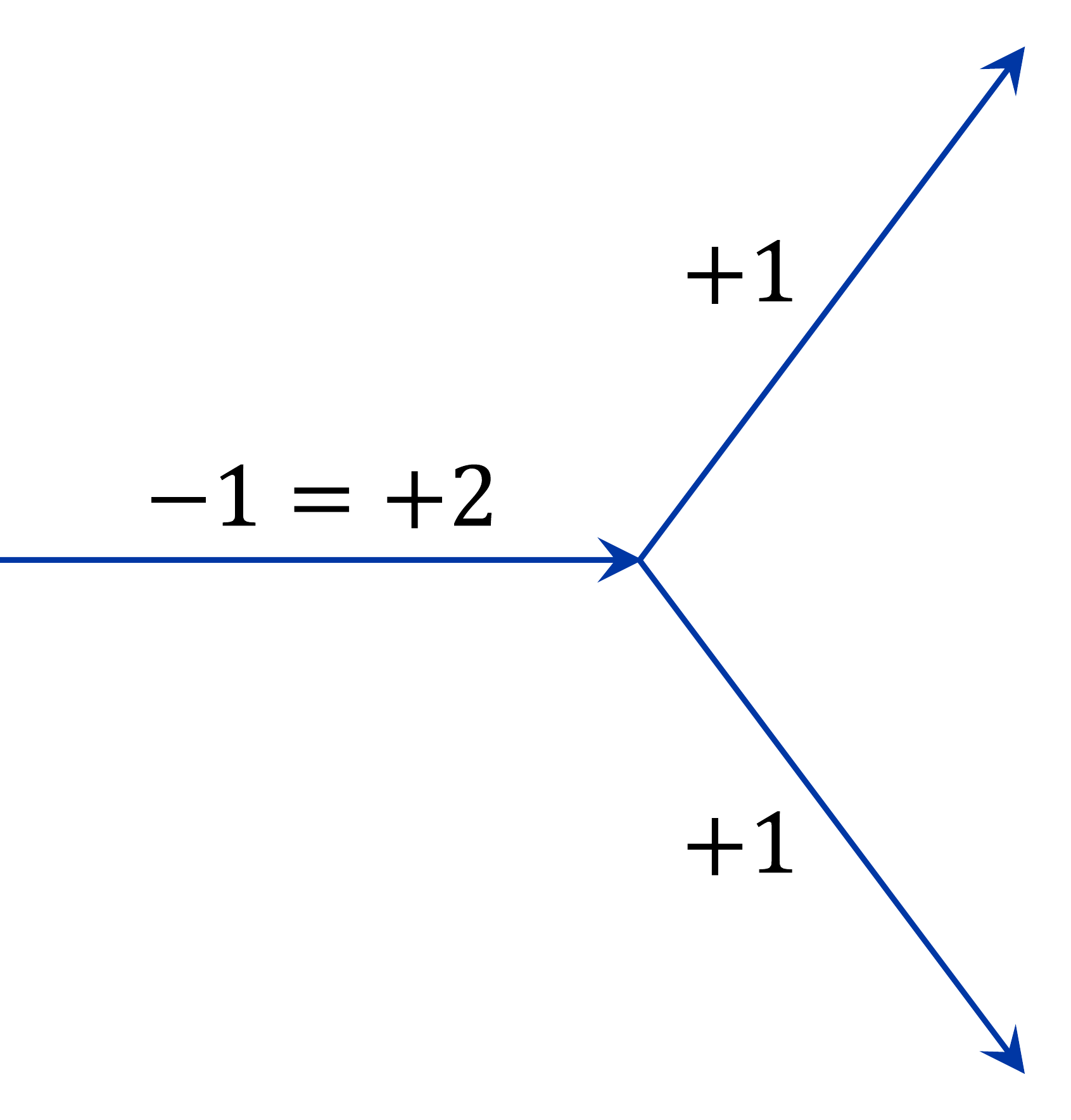}
	
	\vspace{-0.75em}
	
	\caption{\label{fig:branching} Schematic of a monopole vertex (\textbf{left}) versus a branching point (\textbf{right}). The monopole shows the directed flow of \(m = +1\) center charge. Reversal of the left-hand arrow indicates the flow of \(m = -1\) charge, as seen on the right. Due to periodicity in the center charge, \(m = -1\) is equivalent to \(m = +2\). Thus, the right-hand diagram depicts the branching of center charge.}
\end{figure}

Another feature of center-vortex clusters present in \(\mathrm{SU}(N)\) gauge theories for any \(N \geq 3\) is that of vortex branching. In \(\mathrm{SU}(3)\), this involves an \(m = \pm 1\) vortex splitting into two \(m = \mp 1\) vortices. This is allowed due to the conservation of center charge modulo \(N\) in \(\mathrm{SU}(N)\). Since \(m = \pm 1\) center charges simply correspond to opposite orientations in spacetime, branching points can equivalently be interpreted as monopoles in which three vortices of the same center charge emerge from, or converge to, a single point. Branching points and their equivalence to monopoles are illustrated in Fig.~\ref{fig:branching}.

Branching points are inherently three-dimensional objects, and as such are quantified as follows. A three-dimensional slice of the full four-dimensional lattice is taken by holding one dimension \(\mu\) fixed. The branching point density in a given three-dimensional slice is then defined as the proportion of elementary cubes in the slice that contain a branching point,
\begin{equation} \label{eq:branchingpointdensity}
	\rho_\mathrm{branch}(\mu) = \frac{\mathrm{Number\ of\ branchings\ in\ slice}}{N_\mathrm{slice}(\mu)} \,,
\end{equation}
where \(N_\mathrm{slice}(\mu)\) is the three-dimensional volume of the associated slice. This can be averaged over all slices along a given dimension. Similar to the vortex density, \(\rho_\mathrm{branch}\) is usually converted to a physical quantity through dividing by the volume \(a^3\) of an elementary cube, though we again leave it dimensionless.

\begin{figure}
	\includegraphics[width=\linewidth]{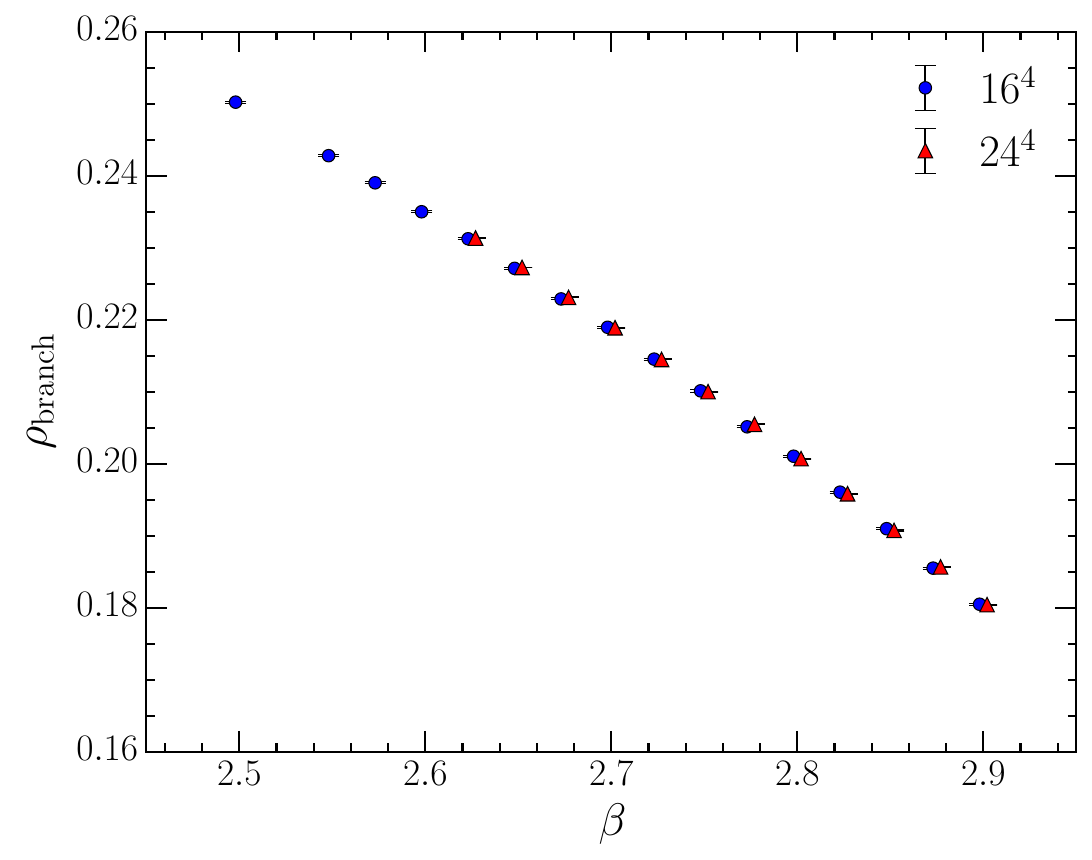}
	
	\vspace{-0.75em}
	
	\caption{\label{fig:branchingpointdensity} The dimensionless branching point density \(\rho_\mathrm{branch}\) defined in Eqs.~(\ref{eq:branchingpointdensity}) and (\ref{eq:branchingpointdensityaveraged}) for each \(\beta\) value and volume. Bootstrap statistical uncertainties are displayed, though are too small to be resolved at this scale. Similar overall behavior to the total vortex density (Fig.~\ref{fig:vortexdensity}) is observed.}
\end{figure}

In Fig.~\ref{fig:branchingpointdensity}, the evolution of a dimensionally averaged branching point density,
\begin{equation} \label{eq:branchingpointdensityaveraged}
	\rho_\mathrm{branch} = \frac{1}{4} \sum_\mu \rho_\mathrm{branch}(\mu) \,,
\end{equation}
is displayed. Here, the same qualitative features are revealed as with the total vortex density in Fig.~\ref{fig:vortexdensity}.
Namely, it exhibits a constant steady rise to the left, attributed to the decreasing \(\beta\) value and growing ``roughness" in the vortex field. As the quantity of center-vortex matter increases, leading to more branching opportunities, one naturally expects to find a greater abundance of branching points in the three-dimensional volume.

\begin{figure}
	\includegraphics[width=\linewidth]{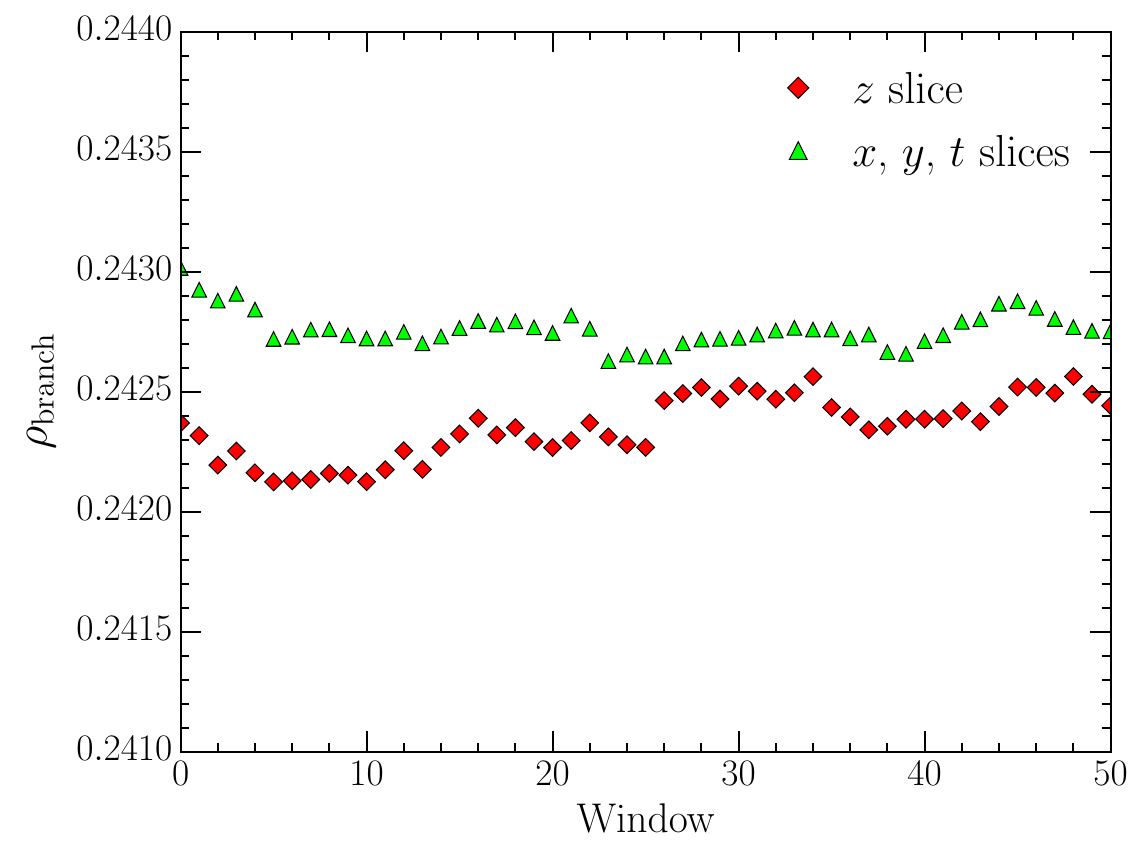}
	
	\vspace{-0.75em}
	
	\caption{\label{fig:dimension_branching_densities} The branching point densities for fixed-\(z\) and combined fixed-\(x, y, t\) slices under a moving average for \(\beta = 2.55\). We find that the branching point density in three-dimensional slices that span the broken \(z\) dimension is persistently larger, in line with the greater density of \(xz\), \(yz\), and \(zt\) plaquettes pierced.}
\end{figure}

We wish to determine whether a similar divergence as in the densities of plaquettes pierced exists for branching points. In this case, the necessary division is into the three-dimensional slices that contain the broken dimension \(\mu\) and the one slice that does not, obtained from slicing through the broken dimension itself (i.e.\ holding the \(\mu\) coordinate fixed). The branching point densities for these two cases are presented as a moving average in Fig.~\ref{fig:dimension_branching_densities}. We use the terminology ``\(\mu\) slice" for the three-dimensional slice obtaining by fixing the \(\mu\) coordinate.

Fig.~\ref{fig:dimension_branching_densities} reveals that there is also a persistent discrepancy between branching points in fixed-\(z\) slices and three-dimensional slices that span the broken \(z\) dimension, with the latter being larger. This follows directly from the greater proportion of pierced plaquettes that span the \(z\) dimension seen in Fig.~\ref{fig:plaquette_densities_averaged}. Even so, the three-dimensional slices that contain the \(z\) dimension are also comprised of plaquettes that do not (e.g.\ fixed-\(t\) slices still contain \(xy\) plaquettes), causing the split in branching point densities to appear even smaller than the corresponding split in vortex densities.

\subsection{Chain lengths} \label{subsec:chainlengths}
So far, the center-vortex properties analyzed have been bulk quantities that average over the entire vortex structure, or three-dimensional slices thereof. We now turn to consider a local aspect of center-vortex geometry, the intrinsic distribution of branching points throughout the vortex cluster. It will be interesting to see whether there is a more substantial difference in such a local quantity compared to the tiny discrepancies identified in the bulk quantities.

This analysis is performed following Ref.~\cite{Biddle:2023lod}. We count the number of vortices between successive branching points along the vortex chain in a three-dimensional slice; we refer to this as the ``chain length" between branching points. A histogram of these chain lengths can then be created, with the \(n\)th bin representing the probability for a vortex to branch after \(n\) steps.

\begin{figure}
	\includegraphics[width=\linewidth]{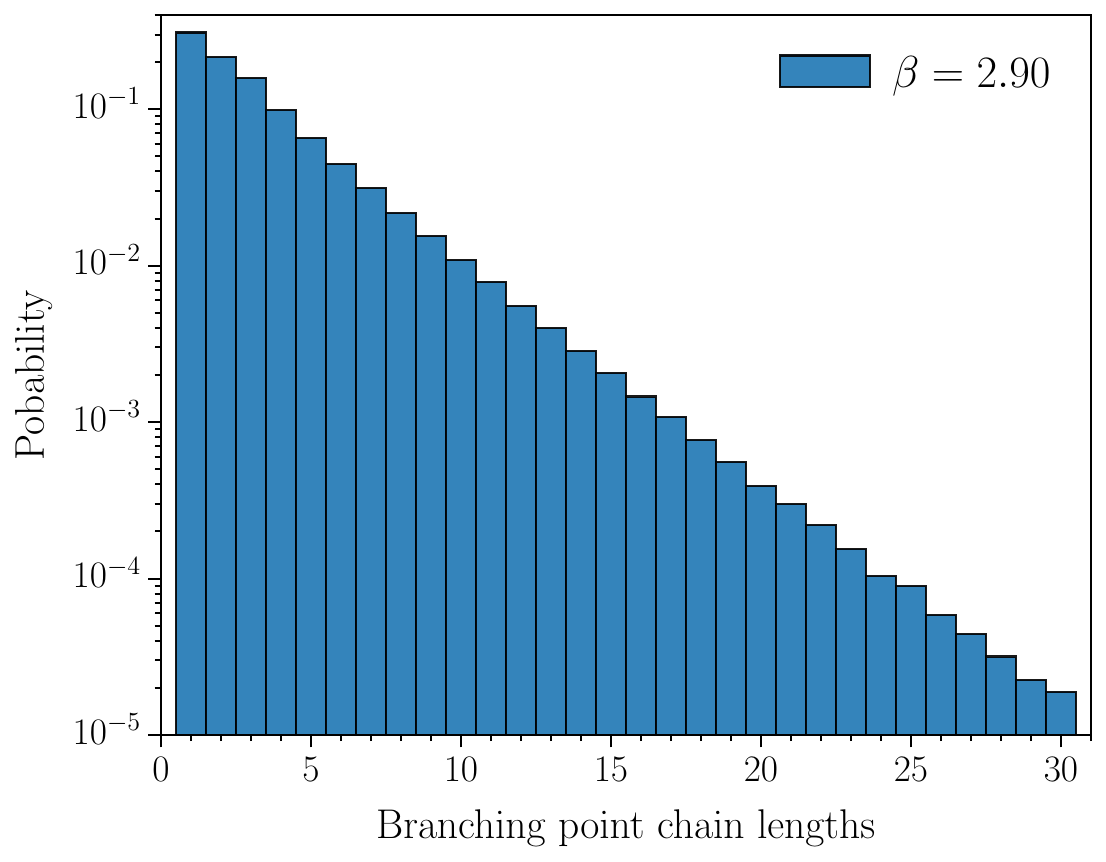}
	
	\vspace{-0.75em}
	
	\caption{\label{fig:b29000chainlengths} A histogram of branching point chain lengths, accumulated over all four slice dimensions, in the physical phase at \(\beta = 2.90\). The histogram is strongly linear on a logarithmic scale, implying the chain lengths are exponentially distributed.}
\end{figure}

An example of such a histogram in the physical phase, at our highest \(\beta\) value of \(\beta = 2.90\), is provided on a logarithmic scale in Fig.~\ref{fig:b29000chainlengths}.
The histogram is seen to be strongly linear, meaning the chain lengths follow a geometric distribution. This has an important consequence, implying there is a constant probability for a vortex to branch at any point along the chain. If this probability is \(p\), then the probability to branch at the \(n\)th step along the chain is
\begin{equation} \label{eq:geometricdistribution}
	P(n) = \left(1 - p\right)^{n-1} \, p \,.
\end{equation}
The geometric distribution can be modelled via an exponential fit of the form
\begin{equation} \label{eq:exponentialdistribution}
	P(n) = \zeta e^{-\beta n} \,,
\end{equation}
with the probability \(p\) related to \(\beta\) via \cite{Biddle:2023lod, Mickley:2024zyg}
\begin{equation}
	p = 1 - e^{-\beta} \,.
\end{equation}
This can be easily seen by equating the ratios of consecutive probabilities \(P(n+1) / P(n)\) in Eqs.~(\ref{eq:geometricdistribution}) and (\ref{eq:exponentialdistribution}). These fits have been performed in past work \cite{Biddle:2023lod, Mickley:2024zyg, Mickley:2024vkm}, though we forgo them here for now as our primary purpose is to identify any change to the chain lengths in the unphysical phase that arises from the asymmetry between plaquettes.

\begin{figure}
	\includegraphics[width=\linewidth]{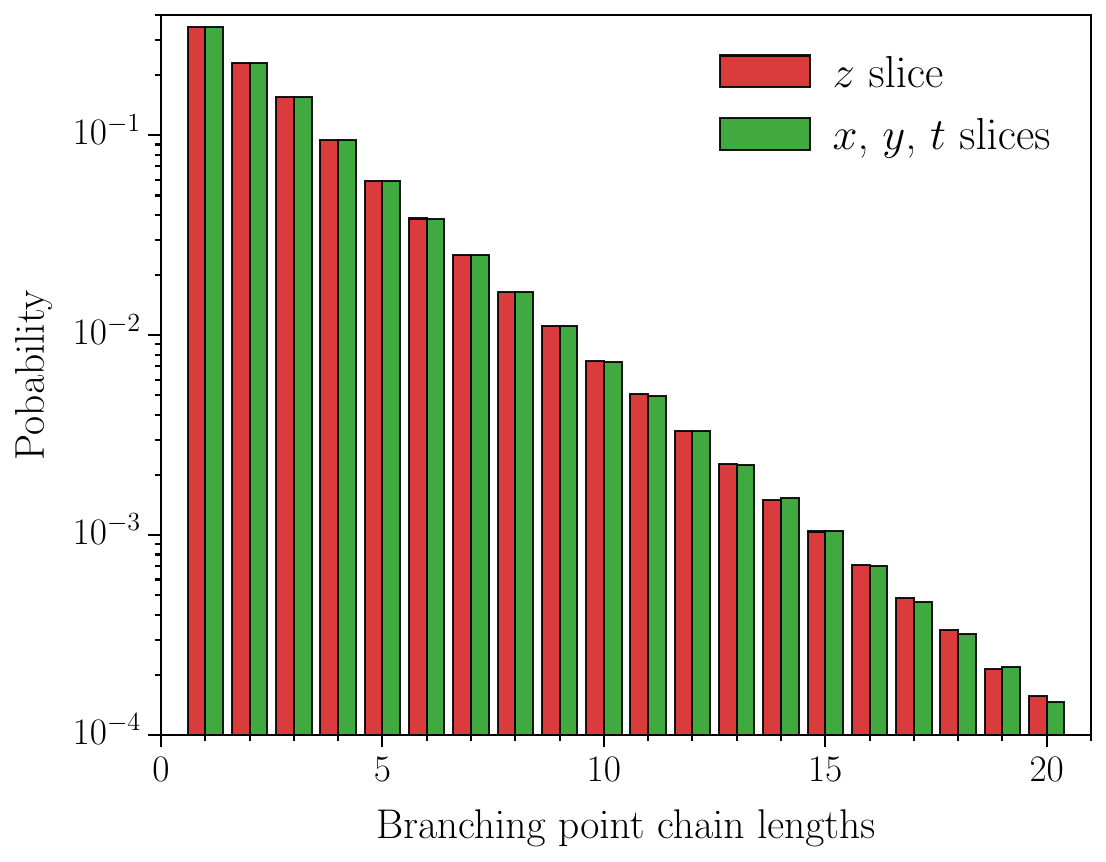}
	
	\vspace{-0.75em}
	
	\caption{\label{fig:b25500chainlengths} Histograms of branching point chain lengths, accumulated separately over fixed-\(z\) slices and the other three slice dimensions, in the unphysical phase at \(\beta = 2.55\). The histograms for these two cases are indistinguishable, such that the broken shift symmetry does not effect any pronounced shift in the chain lengths.}
\end{figure}

To this end, we now produce a chain-length histogram for our example \(\beta = 2.55\) ensemble. As with the branching point density, this is divided into two categories: fixed-\(z\) slices and all other three-dimensional slices, the latter of which span the \(z\) dimension. If a discrepancy were to form, it seems likely that it is the shorter vortex chains that would be impacted. As the chain increases in length, it would become increasingly insensitive to the broken single-site shift symmetry. The histograms for these two cases are presented side-by-side in Fig.~\ref{fig:b25500chainlengths}

Barring small statistical fluctuations at larger chain lengths, we observe no visible difference between the chain-length distributions for fixed-\(z\) slices and the other slice dimensions, not even at short chain lengths. One could imagine that if the probabilities were analyzed at the same level of detail as the vortex and branching point densities, a minute discrepancy would reveal itself. However, the hope by inspecting a local property of vortex structure was to uncover a more pronounced effect of the broken shift symmetry on vortex geometry. From the chain-length histograms in Fig.~\ref{fig:b25500chainlengths}, we conclude that this is not the case.

\section{Conclusion} \label{sec:conclusion}
In this work, we have studied center vortices in the lattice-artefact phase of staggered fermions, in which the single-site shift symmetry of the staggered-fermion action is broken in a finite region of the \((\beta, m)\) phase space. Simulations with six degenerate fermion flavors at \(\beta\) values ranging between \(2.5\)--\(2.9\), and lattice volumes of \(16^4\) and \(24^4\) were employed for the study.

Center vortices were verified to capture the broken shift symmetry manifest in the unphysical phase, as quantified by the order parameter \(\Delta _\mu P\) in Eq.~(\ref{eq:Delta_muP}). This indicated there is a preference for either even or odd plaquettes to be pierced by a vortex in the unphysical phase. The behavior of \(\Delta_\mu P\) along each dimension was examined, demonstrating that the shift symmetry can be broken along multiple dimensions simultaneously and, moreover, that the degree of symmetry breaking along a given dimension can change with simulation time.

The order parameter was further decomposed down to the individual plaquette level, revealing for the first time that only the plaquette orientations spanning the broken dimension are affected. This was established to hold for both the untouched and vortex fields, certifying that center vortices capture the symmetry breaking at the most granular level. Ratios of the number of odd to even plaquettes pierced by a vortex revealed a best-case asymmetry of \(\simeq 10\%\) in our ensembles, making direct observation through visualizations difficult.

Thereafter, various aspects of center-vortex geometry were studied in the physical and unphysical phases. A steady rise in both vortex and branching densities was observed as \(\beta\) decreases. However, no clear transition between the two phases was apparent. Instead, these gentle increases were attributed to a growing roughness in the gauge field as \(\beta\) decreases. This naturally leads to an increase in the quantity of vortex matter, and by extension the density of branching points in the lattice.

A thorough investigation into the vortex density at the individual-plaquette level did reveal a discrepancy that appears in the unphysical phase. Namely, there is a greater proportion of plaquettes pierced for which the shift symmetry is broken (i.e.\ that span the broken dimension). This was seen to extend to the branching point densities, where the same discrepancy was observed in the three-dimensional slices that contain the broken dimension. Still, the effect is small and has a negligible impact on the ``total" vortex and branching point densities that incorporate all plaquettes and slice dimensions, respectively.

Finally, the vortex chain lengths between branching points were analyzed. It was hoped that by examining a more local aspect of vortex geometry, a greater difference arising from the even-odd plaquette asymmetry would reveal itself. Nonetheless, no visible distinction in the chain-length distributions was found, signalling that the even-odd asymmetry has only minor effects on general aspects of center-vortex geometry.

For completeness, we have also examined the vortex correlation measure initially introduced in Ref.~\cite{Mickley:2024zyg} along each dimension in the unphysical phase, for which not even a small difference similar to the vortex and branching point densities was observed.

Perhaps the most important discovery is that the physics of the unphysical broken-shift-symmetry phase is captured by center vortices. This provides a new and relatively simple description of the broken symmetry problem. With further study it holds the promise to reveal a deeper understanding of the physics driving the unphysical phase. The dependence of center-vortex structure on the bare fermion mass \(m\) would also be worth investigating to build up a complete picture of the \((\beta, m)\) phase space around the unphysical phase.

\begin{acknowledgments}
D.~N.\ acknowledges support from the George Southgate Fellowship and thanks the CSSM and Department of Physics at the University of Adelaide for their kind hospitality during his visit where parts of this work were completed. This work was supported with supercomputing resources provided by the Phoenix HPC service at the University of Adelaide. This research was undertaken with the assistance of resources and services from the National Computational Infrastructure (NCI), which is supported by the Australian Government. This research was supported by the Australian Research Council through Grant No.\ DP210103706. D.~N.\ was supported by the NKFIH under grants Excellence No.\ TKP2021-NKTA-64, NKKP Excellence No.\ 151482 and K-147396.
\end{acknowledgments}

\bibliography{main}

\providecommand{\noopsort}[1]{}\providecommand{\singleletter}[1]{#1}%
\begin{thebibliography}{57}%
\makeatletter
\providecommand \@ifxundefined [1]{%
 \@ifx{#1\undefined}
}%
\providecommand \@ifnum [1]{%
 \ifnum #1\expandafter \@firstoftwo
 \else \expandafter \@secondoftwo
 \fi
}%
\providecommand \@ifx [1]{%
 \ifx #1\expandafter \@firstoftwo
 \else \expandafter \@secondoftwo
 \fi
}%
\providecommand \natexlab [1]{#1}%
\providecommand \enquote  [1]{``#1''}%
\providecommand \bibnamefont  [1]{#1}%
\providecommand \bibfnamefont [1]{#1}%
\providecommand \citenamefont [1]{#1}%
\providecommand \href@noop [0]{\@secondoftwo}%
\providecommand \href [0]{\begingroup \@sanitize@url \@href}%
\providecommand \@href[1]{\@@startlink{#1}\@@href}%
\providecommand \@@href[1]{\endgroup#1\@@endlink}%
\providecommand \@sanitize@url [0]{\catcode `\\12\catcode `\$12\catcode
  `\&12\catcode `\#12\catcode `\^12\catcode `\_12\catcode `\%12\relax}%
\providecommand \@@startlink[1]{}%
\providecommand \@@endlink[0]{}%
\providecommand \url  [0]{\begingroup\@sanitize@url \@url }%
\providecommand \@url [1]{\endgroup\@href {#1}{\urlprefix }}%
\providecommand \urlprefix  [0]{URL }%
\providecommand \Eprint [0]{\href }%
\providecommand \doibase [0]{https://doi.org/}%
\providecommand \selectlanguage [0]{\@gobble}%
\providecommand \bibinfo  [0]{\@secondoftwo}%
\providecommand \bibfield  [0]{\@secondoftwo}%
\providecommand \translation [1]{[#1]}%
\providecommand \BibitemOpen [0]{}%
\providecommand \bibitemStop [0]{}%
\providecommand \bibitemNoStop [0]{.\EOS\space}%
\providecommand \EOS [0]{\spacefactor3000\relax}%
\providecommand \BibitemShut  [1]{\csname bibitem#1\endcsname}%
\let\auto@bib@innerbib\@empty
%</preamble>
\bibitem [{\citenamefont {Lee}\ and\ \citenamefont
  {Sharpe}(1999)}]{Lee:1999zxa}%
  \BibitemOpen
  \bibfield  {author} {\bibinfo {author} {\bibfnamefont {W.-J.}\ \bibnamefont
  {Lee}}\ and\ \bibinfo {author} {\bibfnamefont {S.~R.}\ \bibnamefont
  {Sharpe}},\ }\bibfield  {title} {\bibinfo {title} {{Partial flavor symmetry
  restoration for chiral staggered fermions}},\ }\href
  {https://doi.org/10.1103/PhysRevD.60.114503} {\bibfield  {journal} {\bibinfo
  {journal} {Phys. Rev. D}\ }\textbf {\bibinfo {volume} {60}},\ \bibinfo
  {pages} {114503} (\bibinfo {year} {1999})},\ \Eprint
  {https://arxiv.org/abs/hep-lat/9905023} {arXiv:hep-lat/9905023} \BibitemShut
  {NoStop}%
\bibitem [{\citenamefont {Aubin}\ and\ \citenamefont
  {Wang}(2004)}]{Aubin:2004dm}%
  \BibitemOpen
  \bibfield  {author} {\bibinfo {author} {\bibfnamefont {C.}~\bibnamefont
  {Aubin}}\ and\ \bibinfo {author} {\bibfnamefont {Q.-h.}\ \bibnamefont
  {Wang}},\ }\bibfield  {title} {\bibinfo {title} {{Possible Aoki phase for
  staggered fermions}},\ }\href {https://doi.org/10.1103/PhysRevD.70.114504}
  {\bibfield  {journal} {\bibinfo  {journal} {Phys. Rev. D}\ }\textbf {\bibinfo
  {volume} {70}},\ \bibinfo {pages} {114504} (\bibinfo {year} {2004})},\
  \Eprint {https://arxiv.org/abs/hep-lat/0410020} {arXiv:hep-lat/0410020}
  \BibitemShut {NoStop}%
\bibitem [{\citenamefont {Cheng}\ \emph {et~al.}(2012)\citenamefont {Cheng},
  \citenamefont {Hasenfratz},\ and\ \citenamefont {Schaich}}]{Cheng:2011ic}%
  \BibitemOpen
  \bibfield  {author} {\bibinfo {author} {\bibfnamefont {A.}~\bibnamefont
  {Cheng}}, \bibinfo {author} {\bibfnamefont {A.}~\bibnamefont {Hasenfratz}},\
  and\ \bibinfo {author} {\bibfnamefont {D.}~\bibnamefont {Schaich}},\
  }\bibfield  {title} {\bibinfo {title} {{Novel phase in SU(3) lattice gauge
  theory with 12 light fermions}},\ }\href
  {https://doi.org/10.1103/PhysRevD.85.094509} {\bibfield  {journal} {\bibinfo
  {journal} {Phys. Rev. D}\ }\textbf {\bibinfo {volume} {85}},\ \bibinfo
  {pages} {094509} (\bibinfo {year} {2012})},\ \Eprint
  {https://arxiv.org/abs/1111.2317} {arXiv:1111.2317 [hep-lat]} \BibitemShut
  {NoStop}%
\bibitem [{\citenamefont {Hasenfratz}\ \emph {et~al.}(2021)\citenamefont
  {Hasenfratz}, \citenamefont {Shamir},\ and\ \citenamefont
  {Svetitsky}}]{Hasenfratz:2021zsl}%
  \BibitemOpen
  \bibfield  {author} {\bibinfo {author} {\bibfnamefont {A.}~\bibnamefont
  {Hasenfratz}}, \bibinfo {author} {\bibfnamefont {Y.}~\bibnamefont {Shamir}},\
  and\ \bibinfo {author} {\bibfnamefont {B.}~\bibnamefont {Svetitsky}},\
  }\bibfield  {title} {\bibinfo {title} {{Taming lattice artifacts with
  Pauli-Villars fields}},\ }\href {https://doi.org/10.1103/PhysRevD.104.074509}
  {\bibfield  {journal} {\bibinfo  {journal} {Phys. Rev. D}\ }\textbf {\bibinfo
  {volume} {104}},\ \bibinfo {pages} {074509} (\bibinfo {year} {2021})},\
  \Eprint {https://arxiv.org/abs/2109.02790} {arXiv:2109.02790 [hep-lat]}
  \BibitemShut {NoStop}%
\bibitem [{\citenamefont {Kotov}\ \emph {et~al.}(2021)\citenamefont {Kotov},
  \citenamefont {Nogradi}, \citenamefont {Szabo},\ and\ \citenamefont
  {Szikszai}}]{Kotov:2021mgp}%
  \BibitemOpen
  \bibfield  {author} {\bibinfo {author} {\bibfnamefont {A.~Y.}\ \bibnamefont
  {Kotov}}, \bibinfo {author} {\bibfnamefont {D.}~\bibnamefont {Nogradi}},
  \bibinfo {author} {\bibfnamefont {K.~K.}\ \bibnamefont {Szabo}},\ and\
  \bibinfo {author} {\bibfnamefont {L.}~\bibnamefont {Szikszai}},\ }\bibfield
  {title} {\bibinfo {title} {{More on the flavor dependence of $m_\varrho /
  f_\pi$}},\ }\href {https://doi.org/10.1007/JHEP07(2021)202} {\bibfield
  {journal} {\bibinfo  {journal} {J. High Energy Phys.}\ }\textbf {\bibinfo
  {volume} {07}},\ \bibinfo {pages} {202}},\ \bibinfo {note} {[Erratum: J. High
  Energy Phys. 06, 032 (2022)]},\ \Eprint {https://arxiv.org/abs/2107.05996}
  {arXiv:2107.05996 [hep-lat]} \BibitemShut {NoStop}%
\bibitem [{\citenamefont {Aubin}\ \emph {et~al.}(2016)\citenamefont {Aubin},
  \citenamefont {Colletti},\ and\ \citenamefont {Davila}}]{Aubin:2015dgk}%
  \BibitemOpen
  \bibfield  {author} {\bibinfo {author} {\bibfnamefont {C.}~\bibnamefont
  {Aubin}}, \bibinfo {author} {\bibfnamefont {K.}~\bibnamefont {Colletti}},\
  and\ \bibinfo {author} {\bibfnamefont {G.}~\bibnamefont {Davila}},\
  }\bibfield  {title} {\bibinfo {title} {{Unphysical phases in staggered chiral
  perturbation theory}},\ }\href {https://doi.org/10.1103/PhysRevD.93.085009}
  {\bibfield  {journal} {\bibinfo  {journal} {Phys. Rev. D}\ }\textbf {\bibinfo
  {volume} {93}},\ \bibinfo {pages} {085009} (\bibinfo {year} {2016})},\
  \Eprint {https://arxiv.org/abs/1512.01254} {arXiv:1512.01254 [hep-lat]}
  \BibitemShut {NoStop}%
\bibitem [{\citenamefont {Golterman}\ and\ \citenamefont
  {Smit}(1984)}]{Golterman:1984cy}%
  \BibitemOpen
  \bibfield  {author} {\bibinfo {author} {\bibfnamefont {M.~F.~L.}\
  \bibnamefont {Golterman}}\ and\ \bibinfo {author} {\bibfnamefont
  {J.}~\bibnamefont {Smit}},\ }\bibfield  {title} {\bibinfo {title}
  {{Self-energy and flavor interpretation of staggered fermions}},\ }\href
  {https://doi.org/10.1016/0550-3213(84)90424-3} {\bibfield  {journal}
  {\bibinfo  {journal} {Nucl. Phys. B}\ }\textbf {\bibinfo {volume} {245}},\
  \bibinfo {pages} {61} (\bibinfo {year} {1984})}\BibitemShut {NoStop}%
\bibitem [{\citenamefont {'t~Hooft}(1978)}]{tHooft:1977nqb}%
  \BibitemOpen
  \bibfield  {author} {\bibinfo {author} {\bibfnamefont {G.}~\bibnamefont
  {'t~Hooft}},\ }\bibfield  {title} {\bibinfo {title} {{On the phase transition
  towards permanent quark confinement}},\ }\href
  {https://doi.org/10.1016/0550-3213(78)90153-0} {\bibfield  {journal}
  {\bibinfo  {journal} {Nucl. Phys. B}\ }\textbf {\bibinfo {volume} {138}},\
  \bibinfo {pages} {1} (\bibinfo {year} {1978})}\BibitemShut {NoStop}%
\bibitem [{\citenamefont {'t~Hooft}(1979)}]{tHooft:1979rtg}%
  \BibitemOpen
  \bibfield  {author} {\bibinfo {author} {\bibfnamefont {G.}~\bibnamefont
  {'t~Hooft}},\ }\bibfield  {title} {\bibinfo {title} {{A property of electric
  and magnetic flux in non-Abelian gauge theories}},\ }\href
  {https://doi.org/10.1016/0550-3213(79)90595-9} {\bibfield  {journal}
  {\bibinfo  {journal} {Nucl. Phys. B}\ }\textbf {\bibinfo {volume} {153}},\
  \bibinfo {pages} {141} (\bibinfo {year} {1979})}\BibitemShut {NoStop}%
\bibitem [{\citenamefont {Nielsen}\ and\ \citenamefont
  {Olesen}(1979)}]{Nielsen:1979xu}%
  \BibitemOpen
  \bibfield  {author} {\bibinfo {author} {\bibfnamefont {H.~B.}\ \bibnamefont
  {Nielsen}}\ and\ \bibinfo {author} {\bibfnamefont {P.}~\bibnamefont
  {Olesen}},\ }\bibfield  {title} {\bibinfo {title} {{A quantum liquid model
  for the QCD vacuum: Gauge and rotational invariance of domained and quantized
  homogeneous color fields}},\ }\href
  {https://doi.org/10.1016/0550-3213(79)90065-8} {\bibfield  {journal}
  {\bibinfo  {journal} {Nucl. Phys. B}\ }\textbf {\bibinfo {volume} {160}},\
  \bibinfo {pages} {380} (\bibinfo {year} {1979})}\BibitemShut {NoStop}%
\bibitem [{\citenamefont {Greensite}(2003)}]{Greensite:2003bk}%
  \BibitemOpen
  \bibfield  {author} {\bibinfo {author} {\bibfnamefont {J.}~\bibnamefont
  {Greensite}},\ }\bibfield  {title} {\bibinfo {title} {{The confinement
  problem in lattice gauge theory}},\ }\href
  {https://doi.org/10.1016/S0146-6410(03)90012-3} {\bibfield  {journal}
  {\bibinfo  {journal} {Prog. Part. Nucl. Phys.}\ }\textbf {\bibinfo {volume}
  {51}},\ \bibinfo {pages} {1} (\bibinfo {year} {2003})},\ \Eprint
  {https://arxiv.org/abs/hep-lat/0301023} {arXiv:hep-lat/0301023} \BibitemShut
  {NoStop}%
\bibitem [{\citenamefont {Del~Debbio}\ \emph {et~al.}(1997)\citenamefont
  {Del~Debbio}, \citenamefont {Faber}, \citenamefont {Greensite},\ and\
  \citenamefont {Olejn{\'i}k}}]{DelDebbio:1996lih}%
  \BibitemOpen
  \bibfield  {author} {\bibinfo {author} {\bibfnamefont {L.}~\bibnamefont
  {Del~Debbio}}, \bibinfo {author} {\bibfnamefont {M.}~\bibnamefont {Faber}},
  \bibinfo {author} {\bibfnamefont {J.}~\bibnamefont {Greensite}},\ and\
  \bibinfo {author} {\bibfnamefont {{\v S}.}~\bibnamefont {Olejn{\'i}k}},\
  }\bibfield  {title} {\bibinfo {title} {{Center dominance and $Z_2$ vortices
  in SU(2) lattice gauge theory}},\ }\href
  {https://doi.org/10.1103/PhysRevD.55.2298} {\bibfield  {journal} {\bibinfo
  {journal} {Phys. Rev. D}\ }\textbf {\bibinfo {volume} {55}},\ \bibinfo
  {pages} {2298} (\bibinfo {year} {1997})},\ \Eprint
  {https://arxiv.org/abs/hep-lat/9610005} {arXiv:hep-lat/9610005} \BibitemShut
  {NoStop}%
\bibitem [{\citenamefont {Del~Debbio}\ \emph
  {et~al.}(1998{\natexlab{a}})\citenamefont {Del~Debbio}, \citenamefont
  {Faber}, \citenamefont {Greensite},\ and\ \citenamefont
  {Olejn{\'i}k}}]{DelDebbio:1997ep}%
  \BibitemOpen
  \bibfield  {author} {\bibinfo {author} {\bibfnamefont {L.}~\bibnamefont
  {Del~Debbio}}, \bibinfo {author} {\bibfnamefont {M.}~\bibnamefont {Faber}},
  \bibinfo {author} {\bibfnamefont {J.}~\bibnamefont {Greensite}},\ and\
  \bibinfo {author} {\bibfnamefont {{\v S}.}~\bibnamefont {Olejn{\'i}k}},\
  }\bibfield  {title} {\bibinfo {title} {{Center vortices and the asymptotic
  string tension}},\ }\href {https://doi.org/10.1016/S0920-5632(97)00831-1}
  {\bibfield  {journal} {\bibinfo  {journal} {Nucl. Phys. B Proc. Suppl.}\
  }\textbf {\bibinfo {volume} {63}},\ \bibinfo {pages} {552} (\bibinfo {year}
  {1998}{\natexlab{a}})},\ \Eprint {https://arxiv.org/abs/hep-lat/9709032}
  {arXiv:hep-lat/9709032} \BibitemShut {NoStop}%
\bibitem [{\citenamefont {Langfeld}\ \emph {et~al.}(1998)\citenamefont
  {Langfeld}, \citenamefont {Reinhardt},\ and\ \citenamefont
  {Tennert}}]{Langfeld:1997jx}%
  \BibitemOpen
  \bibfield  {author} {\bibinfo {author} {\bibfnamefont {K.}~\bibnamefont
  {Langfeld}}, \bibinfo {author} {\bibfnamefont {H.}~\bibnamefont
  {Reinhardt}},\ and\ \bibinfo {author} {\bibfnamefont {O.}~\bibnamefont
  {Tennert}},\ }\bibfield  {title} {\bibinfo {title} {{Confinement and scaling
  of the vortex vacuum of SU(2) lattice gauge theory}},\ }\href
  {https://doi.org/10.1016/S0370-2693(97)01435-4} {\bibfield  {journal}
  {\bibinfo  {journal} {Phys. Lett. B}\ }\textbf {\bibinfo {volume} {419}},\
  \bibinfo {pages} {317} (\bibinfo {year} {1998})},\ \Eprint
  {https://arxiv.org/abs/hep-lat/9710068} {arXiv:hep-lat/9710068} \BibitemShut
  {NoStop}%
\bibitem [{\citenamefont {Del~Debbio}\ \emph
  {et~al.}(1998{\natexlab{b}})\citenamefont {Del~Debbio}, \citenamefont
  {Faber}, \citenamefont {Giedt}, \citenamefont {Greensite},\ and\
  \citenamefont {Olejn{\'i}k}}]{DelDebbio:1998luz}%
  \BibitemOpen
  \bibfield  {author} {\bibinfo {author} {\bibfnamefont {L.}~\bibnamefont
  {Del~Debbio}}, \bibinfo {author} {\bibfnamefont {M.}~\bibnamefont {Faber}},
  \bibinfo {author} {\bibfnamefont {J.}~\bibnamefont {Giedt}}, \bibinfo
  {author} {\bibfnamefont {J.}~\bibnamefont {Greensite}},\ and\ \bibinfo
  {author} {\bibfnamefont {{\v S}.}~\bibnamefont {Olejn{\'i}k}},\ }\bibfield
  {title} {\bibinfo {title} {{Detection of center vortices in the lattice
  Yang-Mills vacuum}},\ }\href {https://doi.org/10.1103/PhysRevD.58.094501}
  {\bibfield  {journal} {\bibinfo  {journal} {Phys. Rev. D}\ }\textbf {\bibinfo
  {volume} {58}},\ \bibinfo {pages} {094501} (\bibinfo {year}
  {1998}{\natexlab{b}})},\ \Eprint {https://arxiv.org/abs/hep-lat/9801027}
  {arXiv:hep-lat/9801027} \BibitemShut {NoStop}%
\bibitem [{\citenamefont {Faber}\ \emph {et~al.}(1998)\citenamefont {Faber},
  \citenamefont {Greensite},\ and\ \citenamefont {Olejn{\'i}k}}]{Faber:1997rp}%
  \BibitemOpen
  \bibfield  {author} {\bibinfo {author} {\bibfnamefont {M.}~\bibnamefont
  {Faber}}, \bibinfo {author} {\bibfnamefont {J.}~\bibnamefont {Greensite}},\
  and\ \bibinfo {author} {\bibfnamefont {{\v S}.}~\bibnamefont {Olejn{\'i}k}},\
  }\bibfield  {title} {\bibinfo {title} {{Casimir scaling from center vortices:
  Towards an understanding of the adjoint string tension}},\ }\href
  {https://doi.org/10.1103/PhysRevD.57.2603} {\bibfield  {journal} {\bibinfo
  {journal} {Phys. Rev. D}\ }\textbf {\bibinfo {volume} {57}},\ \bibinfo
  {pages} {2603} (\bibinfo {year} {1998})},\ \Eprint
  {https://arxiv.org/abs/hep-lat/9710039} {arXiv:hep-lat/9710039} \BibitemShut
  {NoStop}%
\bibitem [{\citenamefont {Faber}\ \emph
  {et~al.}(1999{\natexlab{a}})\citenamefont {Faber}, \citenamefont
  {Greensite},\ and\ \citenamefont {Olejn{\'i}k}}]{Faber:1998qn}%
  \BibitemOpen
  \bibfield  {author} {\bibinfo {author} {\bibfnamefont {M.}~\bibnamefont
  {Faber}}, \bibinfo {author} {\bibfnamefont {J.}~\bibnamefont {Greensite}},\
  and\ \bibinfo {author} {\bibfnamefont {{\v S}.}~\bibnamefont {Olejn{\'i}k}},\
  }\bibfield  {title} {\bibinfo {title} {{Evidence for a center vortex origin
  of the adjoint string tension}},\ }\href@noop {} {\bibfield  {journal}
  {\bibinfo  {journal} {Acta Phys. Slov.}\ }\textbf {\bibinfo {volume} {49}},\
  \bibinfo {pages} {177} (\bibinfo {year} {1999}{\natexlab{a}})},\ \Eprint
  {https://arxiv.org/abs/hep-lat/9807008} {arXiv:hep-lat/9807008} \BibitemShut
  {NoStop}%
\bibitem [{\citenamefont {Kov{\'a}cs}\ and\ \citenamefont
  {Tomboulis}(1998)}]{Kovacs:1998xm}%
  \BibitemOpen
  \bibfield  {author} {\bibinfo {author} {\bibfnamefont {T.~G.}\ \bibnamefont
  {Kov{\'a}cs}}\ and\ \bibinfo {author} {\bibfnamefont {E.~T.}\ \bibnamefont
  {Tomboulis}},\ }\bibfield  {title} {\bibinfo {title} {{Vortices and
  confinement at weak coupling}},\ }\href
  {https://doi.org/10.1103/PhysRevD.57.4054} {\bibfield  {journal} {\bibinfo
  {journal} {Phys. Rev. D}\ }\textbf {\bibinfo {volume} {57}},\ \bibinfo
  {pages} {4054} (\bibinfo {year} {1998})},\ \Eprint
  {https://arxiv.org/abs/hep-lat/9711009} {arXiv:hep-lat/9711009} \BibitemShut
  {NoStop}%
\bibitem [{\citenamefont {Langfeld}\ \emph {et~al.}(1999)\citenamefont
  {Langfeld}, \citenamefont {Tennert}, \citenamefont {Engelhardt},\ and\
  \citenamefont {Reinhardt}}]{Langfeld:1998cz}%
  \BibitemOpen
  \bibfield  {author} {\bibinfo {author} {\bibfnamefont {K.}~\bibnamefont
  {Langfeld}}, \bibinfo {author} {\bibfnamefont {O.}~\bibnamefont {Tennert}},
  \bibinfo {author} {\bibfnamefont {M.}~\bibnamefont {Engelhardt}},\ and\
  \bibinfo {author} {\bibfnamefont {H.}~\bibnamefont {Reinhardt}},\ }\bibfield
  {title} {\bibinfo {title} {{Center vortices of Yang-Mills theory at finite
  temperatures}},\ }\href {https://doi.org/10.1016/S0370-2693(99)00252-X}
  {\bibfield  {journal} {\bibinfo  {journal} {Phys. Lett. B}\ }\textbf
  {\bibinfo {volume} {452}},\ \bibinfo {pages} {301} (\bibinfo {year}
  {1999})},\ \Eprint {https://arxiv.org/abs/hep-lat/9805002}
  {arXiv:hep-lat/9805002} \BibitemShut {NoStop}%
\bibitem [{\citenamefont {Engelhardt}\ \emph {et~al.}(1998)\citenamefont
  {Engelhardt}, \citenamefont {Langfeld}, \citenamefont {Reinhardt},\ and\
  \citenamefont {Tennert}}]{Engelhardt:1998wu}%
  \BibitemOpen
  \bibfield  {author} {\bibinfo {author} {\bibfnamefont {M.}~\bibnamefont
  {Engelhardt}}, \bibinfo {author} {\bibfnamefont {K.}~\bibnamefont
  {Langfeld}}, \bibinfo {author} {\bibfnamefont {H.}~\bibnamefont
  {Reinhardt}},\ and\ \bibinfo {author} {\bibfnamefont {O.}~\bibnamefont
  {Tennert}},\ }\bibfield  {title} {\bibinfo {title} {{Interaction of confining
  vortices in SU(2) lattice gauge theory}},\ }\href
  {https://doi.org/10.1016/S0370-2693(98)00583-8} {\bibfield  {journal}
  {\bibinfo  {journal} {Phys. Lett. B}\ }\textbf {\bibinfo {volume} {431}},\
  \bibinfo {pages} {141} (\bibinfo {year} {1998})},\ \Eprint
  {https://arxiv.org/abs/hep-lat/9801030} {arXiv:hep-lat/9801030} \BibitemShut
  {NoStop}%
\bibitem [{\citenamefont {Bertle}\ \emph {et~al.}(1999)\citenamefont {Bertle},
  \citenamefont {Faber}, \citenamefont {Greensite},\ and\ \citenamefont
  {Olejn{\'i}k}}]{Bertle:1999tw}%
  \BibitemOpen
  \bibfield  {author} {\bibinfo {author} {\bibfnamefont {R.}~\bibnamefont
  {Bertle}}, \bibinfo {author} {\bibfnamefont {M.}~\bibnamefont {Faber}},
  \bibinfo {author} {\bibfnamefont {J.}~\bibnamefont {Greensite}},\ and\
  \bibinfo {author} {\bibfnamefont {{\v S}.}~\bibnamefont {Olejn{\'i}k}},\
  }\bibfield  {title} {\bibinfo {title} {{The structure of projected center
  vortices in lattice gauge theory}},\ }\href
  {https://doi.org/10.1088/1126-6708/1999/03/019} {\bibfield  {journal}
  {\bibinfo  {journal} {J. High Energy Phys.}\ }\textbf {\bibinfo {volume}
  {03}},\ \bibinfo {pages} {019}},\ \Eprint
  {https://arxiv.org/abs/hep-lat/9903023} {arXiv:hep-lat/9903023} \BibitemShut
  {NoStop}%
\bibitem [{\citenamefont {Engelhardt}\ \emph {et~al.}(2000)\citenamefont
  {Engelhardt}, \citenamefont {Langfeld}, \citenamefont {Reinhardt},\ and\
  \citenamefont {Tennert}}]{Engelhardt:1999fd}%
  \BibitemOpen
  \bibfield  {author} {\bibinfo {author} {\bibfnamefont {M.}~\bibnamefont
  {Engelhardt}}, \bibinfo {author} {\bibfnamefont {K.}~\bibnamefont
  {Langfeld}}, \bibinfo {author} {\bibfnamefont {H.}~\bibnamefont
  {Reinhardt}},\ and\ \bibinfo {author} {\bibfnamefont {O.}~\bibnamefont
  {Tennert}},\ }\bibfield  {title} {\bibinfo {title} {{Deconfinement in SU(2)
  Yang-Mills theory as a center vortex percolation transition}},\ }\href
  {https://doi.org/10.1103/PhysRevD.61.054504} {\bibfield  {journal} {\bibinfo
  {journal} {Phys. Rev. D}\ }\textbf {\bibinfo {volume} {61}},\ \bibinfo
  {pages} {054504} (\bibinfo {year} {2000})},\ \Eprint
  {https://arxiv.org/abs/hep-lat/9904004} {arXiv:hep-lat/9904004} \BibitemShut
  {NoStop}%
\bibitem [{\citenamefont {Engelhardt}\ and\ \citenamefont
  {Reinhardt}(2000)}]{Engelhardt:1999wr}%
  \BibitemOpen
  \bibfield  {author} {\bibinfo {author} {\bibfnamefont {M.}~\bibnamefont
  {Engelhardt}}\ and\ \bibinfo {author} {\bibfnamefont {H.}~\bibnamefont
  {Reinhardt}},\ }\bibfield  {title} {\bibinfo {title} {{Center vortex model
  for the infrared sector of Yang-Mills theory --- confinement and
  deconfinement}},\ }\href {https://doi.org/10.1016/S0550-3213(00)00445-4}
  {\bibfield  {journal} {\bibinfo  {journal} {Nucl. Phys. B}\ }\textbf
  {\bibinfo {volume} {585}},\ \bibinfo {pages} {591} (\bibinfo {year}
  {2000})},\ \Eprint {https://arxiv.org/abs/hep-lat/9912003}
  {arXiv:hep-lat/9912003} \BibitemShut {NoStop}%
\bibitem [{\citenamefont {Faber}\ \emph {et~al.}(2000)\citenamefont {Faber},
  \citenamefont {Greensite},\ and\ \citenamefont {Olejn{\'i}k}}]{Faber:1999sq}%
  \BibitemOpen
  \bibfield  {author} {\bibinfo {author} {\bibfnamefont {M.}~\bibnamefont
  {Faber}}, \bibinfo {author} {\bibfnamefont {J.}~\bibnamefont {Greensite}},\
  and\ \bibinfo {author} {\bibfnamefont {{\v S}.}~\bibnamefont {Olejn{\'i}k}},\
  }\bibfield  {title} {\bibinfo {title} {{First evidence for center dominance
  in SU(3) lattice gauge theory}},\ }\href
  {https://doi.org/10.1016/S0370-2693(00)00013-7} {\bibfield  {journal}
  {\bibinfo  {journal} {Phys. Lett. B}\ }\textbf {\bibinfo {volume} {474}},\
  \bibinfo {pages} {177} (\bibinfo {year} {2000})},\ \Eprint
  {https://arxiv.org/abs/hep-lat/9911006} {arXiv:hep-lat/9911006} \BibitemShut
  {NoStop}%
\bibitem [{\citenamefont {de~Forcrand}\ and\ \citenamefont
  {D'Elia}(1999)}]{deForcrand:1999our}%
  \BibitemOpen
  \bibfield  {author} {\bibinfo {author} {\bibfnamefont {P.}~\bibnamefont
  {de~Forcrand}}\ and\ \bibinfo {author} {\bibfnamefont {M.}~\bibnamefont
  {D'Elia}},\ }\bibfield  {title} {\bibinfo {title} {{Relevance of Center
  Vortices to QCD}},\ }\href {https://doi.org/10.1103/PhysRevLett.82.4582}
  {\bibfield  {journal} {\bibinfo  {journal} {Phys. Rev. Lett.}\ }\textbf
  {\bibinfo {volume} {82}},\ \bibinfo {pages} {4582} (\bibinfo {year}
  {1999})},\ \Eprint {https://arxiv.org/abs/hep-lat/9901020}
  {arXiv:hep-lat/9901020} \BibitemShut {NoStop}%
\bibitem [{\citenamefont {de~Forcrand}\ and\ \citenamefont
  {Pepe}(2001)}]{deForcrand:2000pg}%
  \BibitemOpen
  \bibfield  {author} {\bibinfo {author} {\bibfnamefont {P.}~\bibnamefont
  {de~Forcrand}}\ and\ \bibinfo {author} {\bibfnamefont {M.}~\bibnamefont
  {Pepe}},\ }\bibfield  {title} {\bibinfo {title} {{Center vortices and
  monopoles without lattice Gribov copies}},\ }\href
  {https://doi.org/10.1016/S0550-3213(01)00009-8} {\bibfield  {journal}
  {\bibinfo  {journal} {Nucl. Phys. B}\ }\textbf {\bibinfo {volume} {598}},\
  \bibinfo {pages} {557} (\bibinfo {year} {2001})},\ \Eprint
  {https://arxiv.org/abs/hep-lat/0008016} {arXiv:hep-lat/0008016} \BibitemShut
  {NoStop}%
\bibitem [{\citenamefont {Kov{\'a}cs}\ and\ \citenamefont
  {Tomboulis}(2000)}]{Kovacs:2000sy}%
  \BibitemOpen
  \bibfield  {author} {\bibinfo {author} {\bibfnamefont {T.~G.}\ \bibnamefont
  {Kov{\'a}cs}}\ and\ \bibinfo {author} {\bibfnamefont {E.~T.}\ \bibnamefont
  {Tomboulis}},\ }\bibfield  {title} {\bibinfo {title} {{Computation of the
  Vortex Free Energy in SU(2) Gauge Theory}},\ }\href
  {https://doi.org/10.1103/PhysRevLett.85.704} {\bibfield  {journal} {\bibinfo
  {journal} {Phys. Rev. Lett.}\ }\textbf {\bibinfo {volume} {85}},\ \bibinfo
  {pages} {704} (\bibinfo {year} {2000})},\ \Eprint
  {https://arxiv.org/abs/hep-lat/0002004} {arXiv:hep-lat/0002004} \BibitemShut
  {NoStop}%
\bibitem [{\citenamefont {Langfeld}\ \emph {et~al.}(2002)\citenamefont
  {Langfeld}, \citenamefont {Reinhardt},\ and\ \citenamefont
  {Gattnar}}]{Langfeld:2001cz}%
  \BibitemOpen
  \bibfield  {author} {\bibinfo {author} {\bibfnamefont {K.}~\bibnamefont
  {Langfeld}}, \bibinfo {author} {\bibfnamefont {H.}~\bibnamefont
  {Reinhardt}},\ and\ \bibinfo {author} {\bibfnamefont {J.}~\bibnamefont
  {Gattnar}},\ }\bibfield  {title} {\bibinfo {title} {{Gluon propagator and
  quark confinement}},\ }\href {https://doi.org/10.1016/S0550-3213(01)00574-0}
  {\bibfield  {journal} {\bibinfo  {journal} {Nucl. Phys. B}\ }\textbf
  {\bibinfo {volume} {621}},\ \bibinfo {pages} {131} (\bibinfo {year}
  {2002})},\ \Eprint {https://arxiv.org/abs/hep-ph/0107141}
  {arXiv:hep-ph/0107141} \BibitemShut {NoStop}%
\bibitem [{\citenamefont {Langfeld}(2004)}]{Langfeld:2003ev}%
  \BibitemOpen
  \bibfield  {author} {\bibinfo {author} {\bibfnamefont {K.}~\bibnamefont
  {Langfeld}},\ }\bibfield  {title} {\bibinfo {title} {{Vortex structures in
  pure SU(3) lattice gauge theory}},\ }\href
  {https://doi.org/10.1103/PhysRevD.69.014503} {\bibfield  {journal} {\bibinfo
  {journal} {Phys. Rev. D}\ }\textbf {\bibinfo {volume} {69}},\ \bibinfo
  {pages} {014503} (\bibinfo {year} {2004})},\ \Eprint
  {https://arxiv.org/abs/hep-lat/0307030} {arXiv:hep-lat/0307030} \BibitemShut
  {NoStop}%
\bibitem [{\citenamefont {Engelhardt}\ \emph {et~al.}(2004)\citenamefont
  {Engelhardt}, \citenamefont {Quandt},\ and\ \citenamefont
  {Reinhardt}}]{Engelhardt:2003wm}%
  \BibitemOpen
  \bibfield  {author} {\bibinfo {author} {\bibfnamefont {M.}~\bibnamefont
  {Engelhardt}}, \bibinfo {author} {\bibfnamefont {M.}~\bibnamefont {Quandt}},\
  and\ \bibinfo {author} {\bibfnamefont {H.}~\bibnamefont {Reinhardt}},\
  }\bibfield  {title} {\bibinfo {title} {{Center vortex model for the infrared
  sector of $SU(3)$ Yang-Mills theory---confinement and deconfinement}},\
  }\href {https://doi.org/10.1016/j.nuclphysb.2004.02.036} {\bibfield
  {journal} {\bibinfo  {journal} {Nucl. Phys. B}\ }\textbf {\bibinfo {volume}
  {685}},\ \bibinfo {pages} {227} (\bibinfo {year} {2004})},\ \Eprint
  {https://arxiv.org/abs/hep-lat/0311029} {arXiv:hep-lat/0311029} \BibitemShut
  {NoStop}%
\bibitem [{\citenamefont {Gattnar}\ \emph {et~al.}(2005)\citenamefont
  {Gattnar}, \citenamefont {Gattringer}, \citenamefont {Langfeld},
  \citenamefont {Reinhardt}, \citenamefont {Sch{\"a}fer}, \citenamefont
  {Solbrig},\ and\ \citenamefont {Tok}}]{Gattnar:2004gx}%
  \BibitemOpen
  \bibfield  {author} {\bibinfo {author} {\bibfnamefont {J.}~\bibnamefont
  {Gattnar}}, \bibinfo {author} {\bibfnamefont {C.}~\bibnamefont {Gattringer}},
  \bibinfo {author} {\bibfnamefont {K.}~\bibnamefont {Langfeld}}, \bibinfo
  {author} {\bibfnamefont {H.}~\bibnamefont {Reinhardt}}, \bibinfo {author}
  {\bibfnamefont {A.}~\bibnamefont {Sch{\"a}fer}}, \bibinfo {author}
  {\bibfnamefont {S.}~\bibnamefont {Solbrig}},\ and\ \bibinfo {author}
  {\bibfnamefont {T.}~\bibnamefont {Tok}},\ }\bibfield  {title} {\bibinfo
  {title} {{Center vortices and Dirac eigenmodes in SU(2) lattice gauge
  theory}},\ }\href {https://doi.org/10.1016/j.nuclphysb.2005.03.027}
  {\bibfield  {journal} {\bibinfo  {journal} {Nucl. Phys. B}\ }\textbf
  {\bibinfo {volume} {716}},\ \bibinfo {pages} {105} (\bibinfo {year}
  {2005})},\ \Eprint {https://arxiv.org/abs/hep-lat/0412032}
  {arXiv:hep-lat/0412032} \BibitemShut {NoStop}%
\bibitem [{\citenamefont {Bornyakov}\ \emph {et~al.}(2008)\citenamefont
  {Bornyakov}, \citenamefont {Ilgenfritz}, \citenamefont {Martemyanov},
  \citenamefont {Morozov}, \citenamefont {Muller-Preussker},\ and\
  \citenamefont {Veselov}}]{Bornyakov:2007fz}%
  \BibitemOpen
  \bibfield  {author} {\bibinfo {author} {\bibfnamefont {V.~G.}\ \bibnamefont
  {Bornyakov}}, \bibinfo {author} {\bibfnamefont {E.~M.}\ \bibnamefont
  {Ilgenfritz}}, \bibinfo {author} {\bibfnamefont {B.~V.}\ \bibnamefont
  {Martemyanov}}, \bibinfo {author} {\bibfnamefont {S.~M.}\ \bibnamefont
  {Morozov}}, \bibinfo {author} {\bibfnamefont {M.}~\bibnamefont
  {Muller-Preussker}},\ and\ \bibinfo {author} {\bibfnamefont {A.~I.}\
  \bibnamefont {Veselov}},\ }\bibfield  {title} {\bibinfo {title}
  {{Interrelation between monopoles, vortices, topological charge and chiral
  symmetry breaking: Analysis using overlap fermions for $SU(2)$}},\ }\href
  {https://doi.org/10.1103/PhysRevD.77.074507} {\bibfield  {journal} {\bibinfo
  {journal} {Phys. Rev. D}\ }\textbf {\bibinfo {volume} {77}},\ \bibinfo
  {pages} {074507} (\bibinfo {year} {2008})},\ \Eprint
  {https://arxiv.org/abs/0708.3335} {arXiv:0708.3335 [hep-lat]} \BibitemShut
  {NoStop}%
\bibitem [{\citenamefont {Bowman}\ \emph {et~al.}(2008)\citenamefont {Bowman},
  \citenamefont {Langfeld}, \citenamefont {Leinweber}, \citenamefont {O'~Cais},
  \citenamefont {Sternbeck}, \citenamefont {von Smekal},\ and\ \citenamefont
  {Williams}}]{Bowman:2008qd}%
  \BibitemOpen
  \bibfield  {author} {\bibinfo {author} {\bibfnamefont {P.~O.}\ \bibnamefont
  {Bowman}}, \bibinfo {author} {\bibfnamefont {K.}~\bibnamefont {Langfeld}},
  \bibinfo {author} {\bibfnamefont {D.~B.}\ \bibnamefont {Leinweber}}, \bibinfo
  {author} {\bibfnamefont {A.}~\bibnamefont {O'~Cais}}, \bibinfo {author}
  {\bibfnamefont {A.}~\bibnamefont {Sternbeck}}, \bibinfo {author}
  {\bibfnamefont {L.}~\bibnamefont {von Smekal}},\ and\ \bibinfo {author}
  {\bibfnamefont {A.~G.}\ \bibnamefont {Williams}},\ }\bibfield  {title}
  {\bibinfo {title} {{Center vortices and the quark propagator in SU(2) gauge
  theory}},\ }\href {https://doi.org/10.1103/PhysRevD.78.054509} {\bibfield
  {journal} {\bibinfo  {journal} {Phys. Rev. D}\ }\textbf {\bibinfo {volume}
  {78}},\ \bibinfo {pages} {054509} (\bibinfo {year} {2008})},\ \Eprint
  {https://arxiv.org/abs/0806.4219} {arXiv:0806.4219 [hep-lat]} \BibitemShut
  {NoStop}%
\bibitem [{\citenamefont {Bowman}\ \emph {et~al.}(2011)\citenamefont {Bowman},
  \citenamefont {Langfeld}, \citenamefont {Leinweber}, \citenamefont
  {Sternbeck}, \citenamefont {von Smekal},\ and\ \citenamefont
  {Williams}}]{Bowman:2010zr}%
  \BibitemOpen
  \bibfield  {author} {\bibinfo {author} {\bibfnamefont {P.~O.}\ \bibnamefont
  {Bowman}}, \bibinfo {author} {\bibfnamefont {K.}~\bibnamefont {Langfeld}},
  \bibinfo {author} {\bibfnamefont {D.~B.}\ \bibnamefont {Leinweber}}, \bibinfo
  {author} {\bibfnamefont {A.}~\bibnamefont {Sternbeck}}, \bibinfo {author}
  {\bibfnamefont {L.}~\bibnamefont {von Smekal}},\ and\ \bibinfo {author}
  {\bibfnamefont {A.~G.}\ \bibnamefont {Williams}},\ }\bibfield  {title}
  {\bibinfo {title} {{Role of center vortices in chiral symmetry breaking in
  SU(3) gauge theory}},\ }\href {https://doi.org/10.1103/PhysRevD.84.034501}
  {\bibfield  {journal} {\bibinfo  {journal} {Phys. Rev. D}\ }\textbf {\bibinfo
  {volume} {84}},\ \bibinfo {pages} {034501} (\bibinfo {year} {2011})},\
  \Eprint {https://arxiv.org/abs/1010.4624} {arXiv:1010.4624 [hep-lat]}
  \BibitemShut {NoStop}%
\bibitem [{\citenamefont {O'Malley}\ \emph {et~al.}(2012)\citenamefont
  {O'Malley}, \citenamefont {Kamleh}, \citenamefont {Leinweber},\ and\
  \citenamefont {Moran}}]{OMalley:2011aa}%
  \BibitemOpen
  \bibfield  {author} {\bibinfo {author} {\bibfnamefont {E.-A.}\ \bibnamefont
  {O'Malley}}, \bibinfo {author} {\bibfnamefont {W.}~\bibnamefont {Kamleh}},
  \bibinfo {author} {\bibfnamefont {D.}~\bibnamefont {Leinweber}},\ and\
  \bibinfo {author} {\bibfnamefont {P.}~\bibnamefont {Moran}},\ }\bibfield
  {title} {\bibinfo {title} {{$SU(3)$ centre vortices underpin confinement and
  dynamical chiral symmetry breaking}},\ }\href
  {https://doi.org/10.1103/PhysRevD.86.054503} {\bibfield  {journal} {\bibinfo
  {journal} {Phys. Rev. D}\ }\textbf {\bibinfo {volume} {86}},\ \bibinfo
  {pages} {054503} (\bibinfo {year} {2012})},\ \Eprint
  {https://arxiv.org/abs/1112.2490} {arXiv:1112.2490 [hep-lat]} \BibitemShut
  {NoStop}%
\bibitem [{\citenamefont {H\"ollwieser}\ \emph {et~al.}(2013)\citenamefont
  {H\"ollwieser}, \citenamefont {Schweigler}, \citenamefont {Faber},\ and\
  \citenamefont {Heller}}]{Hollwieser:2013xja}%
  \BibitemOpen
  \bibfield  {author} {\bibinfo {author} {\bibfnamefont {R.}~\bibnamefont
  {H\"ollwieser}}, \bibinfo {author} {\bibfnamefont {T.}~\bibnamefont
  {Schweigler}}, \bibinfo {author} {\bibfnamefont {M.}~\bibnamefont {Faber}},\
  and\ \bibinfo {author} {\bibfnamefont {U.~M.}\ \bibnamefont {Heller}},\
  }\bibfield  {title} {\bibinfo {title} {{Center vortices and chiral symmetry
  breaking in $SU(2)$ lattice gauge theory}},\ }\href
  {https://doi.org/10.1103/PhysRevD.88.114505} {\bibfield  {journal} {\bibinfo
  {journal} {Phys. Rev. D}\ }\textbf {\bibinfo {volume} {88}},\ \bibinfo
  {pages} {114505} (\bibinfo {year} {2013})},\ \Eprint
  {https://arxiv.org/abs/1304.1277} {arXiv:1304.1277 [hep-lat]} \BibitemShut
  {NoStop}%
\bibitem [{\citenamefont {H\"ollwieser}\ \emph {et~al.}(2014)\citenamefont
  {H\"ollwieser}, \citenamefont {Faber}, \citenamefont {Schweigler},\ and\
  \citenamefont {Heller}}]{Hollwieser:2014soz}%
  \BibitemOpen
  \bibfield  {author} {\bibinfo {author} {\bibfnamefont {R.}~\bibnamefont
  {H\"ollwieser}}, \bibinfo {author} {\bibfnamefont {M.}~\bibnamefont {Faber}},
  \bibinfo {author} {\bibfnamefont {T.}~\bibnamefont {Schweigler}},\ and\
  \bibinfo {author} {\bibfnamefont {U.~M.}\ \bibnamefont {Heller}},\ }\bibfield
   {title} {\bibinfo {title} {{Chiral Symmetry Breaking from Center
  Vortices}},\ }\href {https://doi.org/10.22323/1.187.0505} {\bibfield
  {journal} {\bibinfo  {journal} {PoS}\ }\textbf {\bibinfo {volume}
  {LATTICE2013}},\ \bibinfo {pages} {505} (\bibinfo {year} {2014})},\ \Eprint
  {https://arxiv.org/abs/1410.2333} {arXiv:1410.2333 [hep-lat]} \BibitemShut
  {NoStop}%
\bibitem [{\citenamefont {Trewartha}\ \emph {et~al.}(2015)\citenamefont
  {Trewartha}, \citenamefont {Kamleh},\ and\ \citenamefont
  {Leinweber}}]{Trewartha:2015nna}%
  \BibitemOpen
  \bibfield  {author} {\bibinfo {author} {\bibfnamefont {A.}~\bibnamefont
  {Trewartha}}, \bibinfo {author} {\bibfnamefont {W.}~\bibnamefont {Kamleh}},\
  and\ \bibinfo {author} {\bibfnamefont {D.}~\bibnamefont {Leinweber}},\
  }\bibfield  {title} {\bibinfo {title} {{Evidence that centre vortices
  underpin dynamical chiral symmetry breaking in SU(3) gauge theory}},\ }\href
  {https://doi.org/10.1016/j.physletb.2015.06.025} {\bibfield  {journal}
  {\bibinfo  {journal} {Phys. Lett. B}\ }\textbf {\bibinfo {volume} {747}},\
  \bibinfo {pages} {373} (\bibinfo {year} {2015})},\ \Eprint
  {https://arxiv.org/abs/1502.06753} {arXiv:1502.06753 [hep-lat]} \BibitemShut
  {NoStop}%
\bibitem [{\citenamefont {Greensite}(2017)}]{Greensite:2016pfc}%
  \BibitemOpen
  \bibfield  {author} {\bibinfo {author} {\bibfnamefont {J.}~\bibnamefont
  {Greensite}},\ }\bibfield  {title} {\bibinfo {title} {{Confinement from
  Center Vortices: A review of old and new results}},\ }\href
  {https://doi.org/10.1051/epjconf/201713701009} {\bibfield  {journal}
  {\bibinfo  {journal} {EPJ Web Conf.}\ }\textbf {\bibinfo {volume} {137}},\
  \bibinfo {pages} {01009} (\bibinfo {year} {2017})},\ \Eprint
  {https://arxiv.org/abs/1610.06221} {arXiv:1610.06221 [hep-lat]} \BibitemShut
  {NoStop}%
\bibitem [{\citenamefont {Trewartha}\ \emph {et~al.}(2017)\citenamefont
  {Trewartha}, \citenamefont {Kamleh},\ and\ \citenamefont
  {Leinweber}}]{Trewartha:2017ive}%
  \BibitemOpen
  \bibfield  {author} {\bibinfo {author} {\bibfnamefont {A.}~\bibnamefont
  {Trewartha}}, \bibinfo {author} {\bibfnamefont {W.}~\bibnamefont {Kamleh}},\
  and\ \bibinfo {author} {\bibfnamefont {D.}~\bibnamefont {Leinweber}},\
  }\bibfield  {title} {\bibinfo {title} {{Centre vortex removal restores chiral
  symmetry}},\ }\href {https://doi.org/10.1088/1361-6471/aa9443} {\bibfield
  {journal} {\bibinfo  {journal} {J. Phys. G}\ }\textbf {\bibinfo {volume}
  {44}},\ \bibinfo {pages} {125002} (\bibinfo {year} {2017})},\ \Eprint
  {https://arxiv.org/abs/1708.06789} {arXiv:1708.06789 [hep-lat]} \BibitemShut
  {NoStop}%
\bibitem [{\citenamefont {Biddle}\ \emph
  {et~al.}(2022{\natexlab{a}})\citenamefont {Biddle}, \citenamefont {Kamleh},\
  and\ \citenamefont {Leinweber}}]{Biddle:2022zgw}%
  \BibitemOpen
  \bibfield  {author} {\bibinfo {author} {\bibfnamefont {J.~C.}\ \bibnamefont
  {Biddle}}, \bibinfo {author} {\bibfnamefont {W.}~\bibnamefont {Kamleh}},\
  and\ \bibinfo {author} {\bibfnamefont {D.~B.}\ \bibnamefont {Leinweber}},\
  }\bibfield  {title} {\bibinfo {title} {{Static quark potential from center
  vortices in the presence of dynamical fermions}},\ }\href
  {https://doi.org/10.1103/PhysRevD.106.054505} {\bibfield  {journal} {\bibinfo
   {journal} {Phys. Rev. D}\ }\textbf {\bibinfo {volume} {106}},\ \bibinfo
  {pages} {054505} (\bibinfo {year} {2022}{\natexlab{a}})},\ \Eprint
  {https://arxiv.org/abs/2206.00844} {arXiv:2206.00844 [hep-lat]} \BibitemShut
  {NoStop}%
\bibitem [{\citenamefont {Biddle}\ \emph
  {et~al.}(2022{\natexlab{b}})\citenamefont {Biddle}, \citenamefont {Kamleh},\
  and\ \citenamefont {Leinweber}}]{Biddle:2022acd}%
  \BibitemOpen
  \bibfield  {author} {\bibinfo {author} {\bibfnamefont {J.~C.}\ \bibnamefont
  {Biddle}}, \bibinfo {author} {\bibfnamefont {W.}~\bibnamefont {Kamleh}},\
  and\ \bibinfo {author} {\bibfnamefont {D.~B.}\ \bibnamefont {Leinweber}},\
  }\bibfield  {title} {\bibinfo {title} {{Impact of dynamical fermions on the
  center vortex gluon propagator}},\ }\href
  {https://doi.org/10.1103/PhysRevD.106.014506} {\bibfield  {journal} {\bibinfo
   {journal} {Phys. Rev. D}\ }\textbf {\bibinfo {volume} {106}},\ \bibinfo
  {pages} {014506} (\bibinfo {year} {2022}{\natexlab{b}})},\ \Eprint
  {https://arxiv.org/abs/2206.02320} {arXiv:2206.02320 [hep-lat]} \BibitemShut
  {NoStop}%
\bibitem [{\citenamefont {Mickley}\ \emph {et~al.}(2024)\citenamefont
  {Mickley}, \citenamefont {Kamleh},\ and\ \citenamefont
  {Leinweber}}]{Mickley:2024zyg}%
  \BibitemOpen
  \bibfield  {author} {\bibinfo {author} {\bibfnamefont {J.~A.}\ \bibnamefont
  {Mickley}}, \bibinfo {author} {\bibfnamefont {W.}~\bibnamefont {Kamleh}},\
  and\ \bibinfo {author} {\bibfnamefont {D.~B.}\ \bibnamefont {Leinweber}},\
  }\bibfield  {title} {\bibinfo {title} {{Center vortex geometry at finite
  temperature}},\ }\href {https://doi.org/10.1103/PhysRevD.110.034516}
  {\bibfield  {journal} {\bibinfo  {journal} {Phys. Rev. D}\ }\textbf {\bibinfo
  {volume} {110}},\ \bibinfo {pages} {034516} (\bibinfo {year} {2024})},\
  \Eprint {https://arxiv.org/abs/2405.10670} {arXiv:2405.10670 [hep-lat]}
  \BibitemShut {NoStop}%
\bibitem [{\citenamefont {Mickley}\ \emph
  {et~al.}(2025{\natexlab{a}})\citenamefont {Mickley}, \citenamefont {Allton},
  \citenamefont {Bignell},\ and\ \citenamefont {Leinweber}}]{Mickley:2024vkm}%
  \BibitemOpen
  \bibfield  {author} {\bibinfo {author} {\bibfnamefont {J.~A.}\ \bibnamefont
  {Mickley}}, \bibinfo {author} {\bibfnamefont {C.}~\bibnamefont {Allton}},
  \bibinfo {author} {\bibfnamefont {R.}~\bibnamefont {Bignell}},\ and\ \bibinfo
  {author} {\bibfnamefont {D.~B.}\ \bibnamefont {Leinweber}},\ }\bibfield
  {title} {\bibinfo {title} {{Center vortex evidence for a second
  finite-temperature QCD transition}},\ }\href
  {https://doi.org/10.1103/PhysRevD.111.034508} {\bibfield  {journal} {\bibinfo
   {journal} {Phys. Rev. D}\ }\textbf {\bibinfo {volume} {111}},\ \bibinfo
  {pages} {034508} (\bibinfo {year} {2025}{\natexlab{a}})},\ \Eprint
  {https://arxiv.org/abs/2411.19446} {arXiv:2411.19446 [hep-lat]} \BibitemShut
  {NoStop}%
\bibitem [{\citenamefont {Mickley}\ \emph
  {et~al.}(2025{\natexlab{b}})\citenamefont {Mickley}, \citenamefont
  {Leinweber},\ and\ \citenamefont {Nogradi}}]{Mickley:2025mjj}%
  \BibitemOpen
  \bibfield  {author} {\bibinfo {author} {\bibfnamefont {J.~A.}\ \bibnamefont
  {Mickley}}, \bibinfo {author} {\bibfnamefont {D.~B.}\ \bibnamefont
  {Leinweber}},\ and\ \bibinfo {author} {\bibfnamefont {D.}~\bibnamefont
  {Nogradi}},\ }\bibfield  {title} {\bibinfo {title} {{Center vortices and the
  SU(3) conformal window}},\ }\href
  {https://doi.org/10.1103/PhysRevD.111.034516} {\bibfield  {journal} {\bibinfo
   {journal} {Phys. Rev. D}\ }\textbf {\bibinfo {volume} {111}},\ \bibinfo
  {pages} {034516} (\bibinfo {year} {2025}{\natexlab{b}})},\ \Eprint
  {https://arxiv.org/abs/2501.11279} {arXiv:2501.11279 [hep-lat]} \BibitemShut
  {NoStop}%
\bibitem [{\citenamefont {Montero}(1999)}]{Montero:1999by}%
  \BibitemOpen
  \bibfield  {author} {\bibinfo {author} {\bibfnamefont {A.}~\bibnamefont
  {Montero}},\ }\bibfield  {title} {\bibinfo {title} {{Study of SU(3)
  vortex-like configurations with a new maximal center gauge fixing method}},\
  }\href {https://doi.org/10.1016/S0370-2693(99)01113-2} {\bibfield  {journal}
  {\bibinfo  {journal} {Phys. Lett. B}\ }\textbf {\bibinfo {volume} {467}},\
  \bibinfo {pages} {106} (\bibinfo {year} {1999})},\ \Eprint
  {https://arxiv.org/abs/hep-lat/9906010} {arXiv:hep-lat/9906010} \BibitemShut
  {NoStop}%
\bibitem [{\citenamefont {Faber}\ \emph
  {et~al.}(1999{\natexlab{b}})\citenamefont {Faber}, \citenamefont {Greensite},
  \citenamefont {Olejn{\'i}k},\ and\ \citenamefont {Yamada}}]{Faber:1999gu}%
  \BibitemOpen
  \bibfield  {author} {\bibinfo {author} {\bibfnamefont {M.}~\bibnamefont
  {Faber}}, \bibinfo {author} {\bibfnamefont {J.}~\bibnamefont {Greensite}},
  \bibinfo {author} {\bibfnamefont {{\v S}.}~\bibnamefont {Olejn{\'i}k}},\ and\
  \bibinfo {author} {\bibfnamefont {D.}~\bibnamefont {Yamada}},\ }\bibfield
  {title} {\bibinfo {title} {{The vortex-finding property of maximal center
  (and other) gauges}},\ }\href {https://doi.org/10.1088/1126-6708/1999/12/012}
  {\bibfield  {journal} {\bibinfo  {journal} {J. High Energy Phys.}\ }\textbf
  {\bibinfo {volume} {12}},\ \bibinfo {pages} {012}},\ \Eprint
  {https://arxiv.org/abs/hep-lat/9910033} {arXiv:hep-lat/9910033} \BibitemShut
  {NoStop}%
\bibitem [{\citenamefont {Duane}\ \emph {et~al.}(1987)\citenamefont {Duane},
  \citenamefont {Kennedy}, \citenamefont {Pendleton},\ and\ \citenamefont
  {Roweth}}]{Duane:1987de}%
  \BibitemOpen
  \bibfield  {author} {\bibinfo {author} {\bibfnamefont {S.}~\bibnamefont
  {Duane}}, \bibinfo {author} {\bibfnamefont {A.~D.}\ \bibnamefont {Kennedy}},
  \bibinfo {author} {\bibfnamefont {B.~J.}\ \bibnamefont {Pendleton}},\ and\
  \bibinfo {author} {\bibfnamefont {D.}~\bibnamefont {Roweth}},\ }\bibfield
  {title} {\bibinfo {title} {{Hybrid Monte Carlo}},\ }\href
  {https://doi.org/10.1016/0370-2693(87)91197-X} {\bibfield  {journal}
  {\bibinfo  {journal} {Phys. Lett. B}\ }\textbf {\bibinfo {volume} {195}},\
  \bibinfo {pages} {216} (\bibinfo {year} {1987})}\BibitemShut {NoStop}%
\bibitem [{\citenamefont {Clark}\ and\ \citenamefont
  {Kennedy}(2007)}]{Clark:2006fx}%
  \BibitemOpen
  \bibfield  {author} {\bibinfo {author} {\bibfnamefont {M.~A.}\ \bibnamefont
  {Clark}}\ and\ \bibinfo {author} {\bibfnamefont {A.~D.}\ \bibnamefont
  {Kennedy}},\ }\bibfield  {title} {\bibinfo {title} {{Accelerating
  Dynamical-Fermion Computations Using the Rational Hybrid Monte Carlo
  Algorithm with Multiple Pseudofermion Fields}},\ }\href
  {https://doi.org/10.1103/PhysRevLett.98.051601} {\bibfield  {journal}
  {\bibinfo  {journal} {Phys. Rev. Lett.}\ }\textbf {\bibinfo {volume} {98}},\
  \bibinfo {pages} {051601} (\bibinfo {year} {2007})},\ \Eprint
  {https://arxiv.org/abs/hep-lat/0608015} {arXiv:hep-lat/0608015} \BibitemShut
  {NoStop}%
\bibitem [{\citenamefont {Morningstar}\ and\ \citenamefont
  {Peardon}(2004)}]{Morningstar:2003gk}%
  \BibitemOpen
  \bibfield  {author} {\bibinfo {author} {\bibfnamefont {C.}~\bibnamefont
  {Morningstar}}\ and\ \bibinfo {author} {\bibfnamefont {M.~J.}\ \bibnamefont
  {Peardon}},\ }\bibfield  {title} {\bibinfo {title} {{Analytic smearing of
  SU(3) link variables in lattice QCD}},\ }\href
  {https://doi.org/10.1103/PhysRevD.69.054501} {\bibfield  {journal} {\bibinfo
  {journal} {Phys. Rev. D}\ }\textbf {\bibinfo {volume} {69}},\ \bibinfo
  {pages} {054501} (\bibinfo {year} {2004})},\ \Eprint
  {https://arxiv.org/abs/hep-lat/0311018} {arXiv:hep-lat/0311018} \BibitemShut
  {NoStop}%
\bibitem [{\citenamefont {Durr}\ \emph {et~al.}(2011)\citenamefont {Durr},
  \citenamefont {Fodor}, \citenamefont {Hoelbling}, \citenamefont {Katz},
  \citenamefont {Krieg}, \citenamefont {Kurth}, \citenamefont {Lellouch},
  \citenamefont {Lippert}, \citenamefont {Szabo},\ and\ \citenamefont
  {Vulvert}}]{BMW:2010skj}%
  \BibitemOpen
  \bibfield  {author} {\bibinfo {author} {\bibfnamefont {S.}~\bibnamefont
  {Durr}}, \bibinfo {author} {\bibfnamefont {Z.}~\bibnamefont {Fodor}},
  \bibinfo {author} {\bibfnamefont {C.}~\bibnamefont {Hoelbling}}, \bibinfo
  {author} {\bibfnamefont {S.~D.}\ \bibnamefont {Katz}}, \bibinfo {author}
  {\bibfnamefont {S.}~\bibnamefont {Krieg}}, \bibinfo {author} {\bibfnamefont
  {T.}~\bibnamefont {Kurth}}, \bibinfo {author} {\bibfnamefont
  {L.}~\bibnamefont {Lellouch}}, \bibinfo {author} {\bibfnamefont
  {T.}~\bibnamefont {Lippert}}, \bibinfo {author} {\bibfnamefont {K.~K.}\
  \bibnamefont {Szabo}},\ and\ \bibinfo {author} {\bibfnamefont
  {G.}~\bibnamefont {Vulvert}} (\bibinfo {collaboration} {BMW}),\ }\bibfield
  {title} {\bibinfo {title} {{Lattice QCD at the physical point: Simulation and
  analysis details}},\ }\href {https://doi.org/10.1007/JHEP08(2011)148}
  {\bibfield  {journal} {\bibinfo  {journal} {J. High Energy Phys.}\ }\textbf
  {\bibinfo {volume} {08}},\ \bibinfo {pages} {148}},\ \Eprint
  {https://arxiv.org/abs/1011.2711} {arXiv:1011.2711 [hep-lat]} \BibitemShut
  {NoStop}%
\bibitem [{\citenamefont {Symanzik}(1983{\natexlab{a}})}]{Symanzik:1983dc}%
  \BibitemOpen
  \bibfield  {author} {\bibinfo {author} {\bibfnamefont {K.}~\bibnamefont
  {Symanzik}},\ }\bibfield  {title} {\bibinfo {title} {{Continuum limit and
  improved action in lattice theories: (I). Principles and $\phi^4$ theory}},\
  }\href {https://doi.org/10.1016/0550-3213(83)90468-6} {\bibfield  {journal}
  {\bibinfo  {journal} {Nucl. Phys.}\ }\textbf {\bibinfo {volume} {B226}},\
  \bibinfo {pages} {187} (\bibinfo {year} {1983}{\natexlab{a}})}\BibitemShut
  {NoStop}%
%%CITATION = NUPHA,B226,187;%%
\bibitem [{\citenamefont {Symanzik}(1983{\natexlab{b}})}]{Symanzik:1983gh}%
  \BibitemOpen
  \bibfield  {author} {\bibinfo {author} {\bibfnamefont {K.}~\bibnamefont
  {Symanzik}},\ }\bibfield  {title} {\bibinfo {title} {{Continuum limit and
  improved action in lattice theories: (II). $O(N)$ non-linear sigma model in
  perturbation theory}},\ }\href {https://doi.org/10.1016/0550-3213(83)90469-8}
  {\bibfield  {journal} {\bibinfo  {journal} {Nucl. Phys.}\ }\textbf {\bibinfo
  {volume} {B226}},\ \bibinfo {pages} {205} (\bibinfo {year}
  {1983}{\natexlab{b}})}\BibitemShut {NoStop}%
%%CITATION = NUPHA,B226,205;%%
\bibitem [{\citenamefont {Nogradi}\ \emph {et~al.}(2019)\citenamefont
  {Nogradi}, \citenamefont {Nogradi}, \citenamefont {Szikszai},\ and\
  \citenamefont {Szikszai}}]{Nogradi:2019iek}%
  \BibitemOpen
  \bibfield  {author} {\bibinfo {author} {\bibfnamefont {D.}~\bibnamefont
  {Nogradi}}, \bibinfo {author} {\bibfnamefont {D.}~\bibnamefont {Nogradi}},
  \bibinfo {author} {\bibfnamefont {L.}~\bibnamefont {Szikszai}},\ and\
  \bibinfo {author} {\bibfnamefont {L.}~\bibnamefont {Szikszai}},\ }\bibfield
  {title} {\bibinfo {title} {{The flavor dependence of $m_\varrho / f_\pi$}},\
  }\href {https://doi.org/10.1007/JHEP05(2019)197} {\bibfield  {journal}
  {\bibinfo  {journal} {J. High Energy Phys.}\ }\textbf {\bibinfo {volume}
  {05}},\ \bibinfo {pages} {197}},\ \bibinfo {note} {[Erratum: J. High Energy
  Phys. 06, 031 (2022)]},\ \Eprint {https://arxiv.org/abs/1905.01909}
  {arXiv:1905.01909 [hep-lat]} \BibitemShut {NoStop}%
\bibitem [{\citenamefont {Nogradi}\ and\ \citenamefont
  {Szikszai}(2019)}]{Nogradi:2019auv}%
  \BibitemOpen
  \bibfield  {author} {\bibinfo {author} {\bibfnamefont {D.}~\bibnamefont
  {Nogradi}}\ and\ \bibinfo {author} {\bibfnamefont {L.}~\bibnamefont
  {Szikszai}},\ }\bibfield  {title} {\bibinfo {title} {{The model dependence of
  $m_\varrho / f_\pi$}},\ }\href {https://doi.org/10.22323/1.363.0237}
  {\bibfield  {journal} {\bibinfo  {journal} {PoS}\ }\textbf {\bibinfo {volume}
  {LATTICE2019}},\ \bibinfo {pages} {237} (\bibinfo {year} {2019})},\ \Eprint
  {https://arxiv.org/abs/1912.04114} {arXiv:1912.04114 [hep-lat]} \BibitemShut
  {NoStop}%
\bibitem [{\citenamefont {Biddle}\ \emph {et~al.}(2020)\citenamefont {Biddle},
  \citenamefont {Kamleh},\ and\ \citenamefont {Leinweber}}]{Biddle:2019gke}%
  \BibitemOpen
  \bibfield  {author} {\bibinfo {author} {\bibfnamefont {J.~C.}\ \bibnamefont
  {Biddle}}, \bibinfo {author} {\bibfnamefont {W.}~\bibnamefont {Kamleh}},\
  and\ \bibinfo {author} {\bibfnamefont {D.~B.}\ \bibnamefont {Leinweber}},\
  }\bibfield  {title} {\bibinfo {title} {{Visualization of center vortex
  structure}},\ }\href {https://doi.org/10.1103/PhysRevD.102.034504} {\bibfield
   {journal} {\bibinfo  {journal} {Phys. Rev. D}\ }\textbf {\bibinfo {volume}
  {102}},\ \bibinfo {pages} {034504} (\bibinfo {year} {2020})},\ \Eprint
  {https://arxiv.org/abs/1912.09531} {arXiv:1912.09531 [hep-lat]} \BibitemShut
  {NoStop}%
\bibitem [{\citenamefont {Biddle}\ \emph {et~al.}(2023)\citenamefont {Biddle},
  \citenamefont {Kamleh},\ and\ \citenamefont {Leinweber}}]{Biddle:2023lod}%
  \BibitemOpen
  \bibfield  {author} {\bibinfo {author} {\bibfnamefont {J.~C.}\ \bibnamefont
  {Biddle}}, \bibinfo {author} {\bibfnamefont {W.}~\bibnamefont {Kamleh}},\
  and\ \bibinfo {author} {\bibfnamefont {D.~B.}\ \bibnamefont {Leinweber}},\
  }\bibfield  {title} {\bibinfo {title} {{Center vortex structure in the
  presence of dynamical fermions}},\ }\href
  {https://doi.org/10.1103/PhysRevD.107.094507} {\bibfield  {journal} {\bibinfo
   {journal} {Phys. Rev. D}\ }\textbf {\bibinfo {volume} {107}},\ \bibinfo
  {pages} {094507} (\bibinfo {year} {2023})},\ \Eprint
  {https://arxiv.org/abs/2302.05897} {arXiv:2302.05897 [hep-lat]} \BibitemShut
  {NoStop}%
\end{thebibliography}%

\end{document}